\documentclass[12pt]{article}
\pdfoutput=1
\usepackage{amsmath,amsfonts,graphicx,color,bbm,tikz}
\usepackage{comment}
\usepackage[nosort]{cite}
\usepackage{subfigure}
\usetikzlibrary{calc,positioning}
\usetikzlibrary{patterns,arrows,decorations.pathreplacing}
\usepackage{caption}
\usepackage{ulem}
\tikzset{>=stealth}

\makeatletter\@addtoreset{equation}{section}\makeatother

\newcommand{\be}{\begin{equation}}
\newcommand{\ee}{\end{equation}}
\newcommand{\bea}{\begin{eqnarray}}
\newcommand{\eea}{\end{eqnarray}}
\newcommand{\Tr}{{\rm Tr\,}}

\newcommand{\cS}{{\cal S}}

\newcommand{\cO}{{\cal O}}

\newcommand{\bra}[1]{{\left< {#1} \right|}}
\newcommand{\ket}[1]{{\left| {#1} \right>}}

\def\nn{\nonumber}

\newcommand{\binomi}[2]{\begin{pmatrix} #1 \\ #2 \end{pmatrix}}

\renewcommand{\title}[1]{\vbox{\center\LARGE{#1}}\vspace{3mm}}
\renewcommand{\author}[1]{\vbox{\center#1}\vspace{3mm}}

\newcommand{\email}[1]{\vbox{\center\tt#1}\vspace{3mm}}

\begin{document}

\rightline{\small{\tt }}
\begin{center}

\vskip-1.5cm
{\large {\bf Quantum Phase Transitions and Localization in Semigroup Fredkin Spin Chain} }
\vskip 0.75cm

Pramod Padmanabhan$^*$, Fumihiko Sugino$^*$ and Vladimir Korepin$^\dagger$, 

\vskip 0.5cm 
${}^*$ Fields, Gravity \& Strings, Center for Theoretical Physics of the Universe,\\
Institute for Basic Science (IBS), 55, Expo-ro, Yuseong-gu, Daejeon 34126, Republic of Korea\\
${}^\dagger $ C.N.Yang Institute for Theoretical Physics, Stony Brook University, NY 11794, USA\\
\vskip 0.5cm 

\email{pramod23phys@gmail.com, fusugino@gmail.com, korepin@gmail.com}

\vskip 0.5cm 

\end{center}


\abstract{
\noindent 
We construct an extended quantum spin chain model by introducing new degrees of freedom to the Fredkin spin chain. 
The new degrees of freedom called arrow indices are partly associated to the symmetric inverse semigroup $\cS^3_1$. Ground states of the model 
fall into three different phases, and quantum phase transition takes place at each phase boundary. One of the phases exhibits logarithmic violation of 
the area law of entanglement entropy and quantum criticality, whereas the other two obey the area law. 
As an interesting feature arising by the extension, there are excited states due to disconnections with respect to the arrow indices. 
We show that these states are localized without disorder. 
}


\section{Introduction}
Symmetry generated by a group acts as a guiding principle to construct models. 
As an extension of this, we construct a model based on an action of a semigroup. Inverse semigroups are discussed as symmetries 
of the tilings of $\mathbb{R}^n$ and aperiodic structures like quasicrystals~\cite{3p,6p,8p}. Symmetric inverse semigroups (SISs) that are analogous to the permutation group $S_n$ in the ordinary group  
is applied in constructing integrable supersymmetric many-body systems~\cite{PP}.  
 
In the previous paper~\cite{FSPP}, we have discussed extensions of the Motzkin spin chain~\cite{bravyi,shor} based on the SISs $\cS^2_1, \cS^3_1, \cS^3_2$. 
In this paper, an analogous extension is made for the Fredkin spin chain~\cite{dellanna,dyck} based on the SIS $\cS^3_1$. 
The original Fredkin model has an unique ground state, which corresponds to certain random paths on a 2D plane, called  {\it Dyck walks} (DWs).   
The modification introduces arrow indices (partly associated to $\cS^3_1$) to all the steps of the DWs, which yields the ground state degeneracy (GSD). 
In addition, we can introduce tunable parameters while maintaining the frustration-free nature of the Hamiltonian. 
For ground states, we find three phases in the parameter space. One of the phases exhibits a logarithmic violation of the area law of the entanglement entropy (EE),  
which indicates quantum criticality. 
On the other hand, the remaining two phases provide area law behavior for the EE. 
As an interesting feature of this extension which has no analog in the original model, there are excited states corresponding to disconnected paths in DWs with arrow indices. 
For such states, any connected two-point correlation function of local operators vanishes, when the two local operators act on states corresponding to separate connected components of the paths. 
This implies that information is confined to each of the connected components indicating localization. 
In contrast to the ordinary case of localization~\cite{anderson,basko,MBLrev1,MBLrev2,MBLrev3}, the localization in our system occurs without introducing a random noise. 

The paper is organized as follows. In the next section, we construct the extended model of the Fredkin spin chain based on the SIS $\cS^3_1$. In section~\ref{sec:gs}, 
we investigate the structure of ground states and their degeneracies for the three phases. In section~\ref{sec:EEGS}, EEs of the ground states are computed, 
and quantum phase transitions are shown. 
In section~\ref{sec:excitations}, we discuss excited states corresponding to disconnected paths in DWs with arrow indices, and show that they are localized. 
Section~\ref{sec:discussions} is devoted to summary of the result, announcement of the forthcoming paper~\cite{PPFSVK2} and possible future directions.

\section{The modified Fredkin chain}
\label{sec:mFredkin}
The local Hilbert space of the Fredkin spin chain consists of $\{\ket{\uparrow}, \ket{\downarrow}\}$ states. 
These states are mapped to ``up'' and ``down'' arrows in the 2D plane, pointing along $(1, 1)$ and $(1,-1)$ respectively. 
With this interpretation, the states in the global Hilbert space, constructed on a 1D chain of the tensor products of the local Hilbert spaces, 
can be thought of as paths on the 2D plane. A particular set of such paths, called DWs, starts at the origin 
and ends on the positive $x$-axis staying in the positive quadrant all along the walk. The sum of such paths forms the ground state of the Fredkin spin chain.

\paragraph {The new Hilbert space} -  
In the modified Fredkin chain, we replace the above local Hilbert space with $\{\ket{x_{1,2}}, \ket{x_{1,3}}, \ket{x_{2,3}}, \ket{x_{2,1}}, \ket{x_{3,1}}, \ket{x_{3,2}}\}$. 
The states  $\ket{x_{a,b}}$ represent up arrows when $a<b$ and down arrows when $a>b$. 
Thus we have three ways to move up using $\ket{x_{1,2}}$, $\ket{x_{1,3}}$ and $\ket{x_{2,3}}$ 
and three ways to move down $\ket{x_{2,1}}$, $\ket{x_{3,1}}$ and $\ket{x_{3,2}}$. 
This modification is partly associated to the SIS $\cS^3_1$~\footnote{
Note that if we include the elements $\ket{x_{1,1}}$, $\ket{x_{2,2}}$ and $\ket{x_{3,3}}$ 
then the new basis of 9 elements carries the representation of the SIS, $\cS^3_1$. 
This is the situation in the Motzkin spin chain where $\ket{x_{a,a}}$ represent the flat arrows \cite{FSPP}. 
However in the Fredkin case we do not have flat arrows and the remaining six elements do not carry the representation of $\cS^3_1$. 
Hence we avoid calling this spin chain the $\cS^3_1$-Fredkin chain. However they do carry the representation of $\cS^2_1\oplus \cS^2_1\oplus \cS^2_1$
}. 

\begin{figure}[h!]
\captionsetup{width=0.8\textwidth}
\begin{center}
		\includegraphics[scale=0.8]{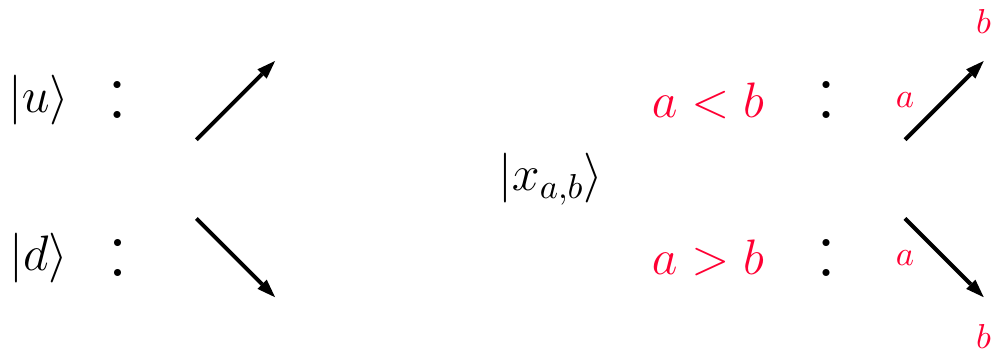} 
	\caption{\small Hilbert space for the modified Fredkin chain.  
	 }
\label{hilb}
\end{center}
\end{figure}

The modification can be understood as introducing an additional degree of freedom, called the {\it arrow index}, to the two ends of the arrow (In the modified Motzkin case~\cite{FSPP}, we called them the semigroup index). 
That is for the basis state $\ket{x_{a,b}}$, $a$ and $b$ denote the arrow indices with $a, b\in \{1,2,3\}$. 
On the other hand, in the colored case of the modified Fredkin chain, apart from the arrow indices, we also include a color degree of freedom to the arrow of each basis state~\cite{PPFSVK2}. 
The Hilbert spaces are illustrated in Fig.~\ref{hilb}. Henceforth we will use the words paths and states to mean the same thing.

The introduction of the arrow indices brings about two important changes in the paths of the modified DWs: different kinds of paths and a change in the maximum heights reached in a given path. 

\begin{figure}[h!]
\captionsetup{width=0.8\textwidth}
\begin{center}
		\includegraphics[scale=0.8]{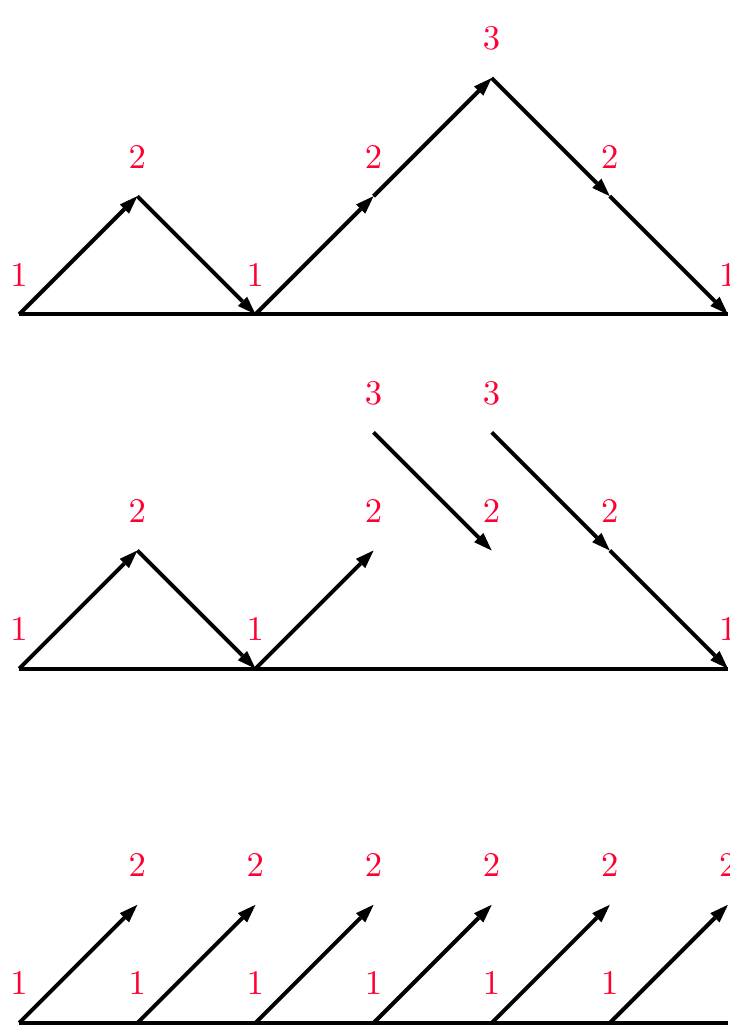} 
	\caption{\small The different kinds of paths possible in the modified Fredkin system. The figure shows a fully connected, a partially connected and a totally disconnected path from top to bottom on a 6-link path.}
\label{dpaths}
\end{center}
\end{figure}

\paragraph {The different kinds of paths} - 
The paths now split as {\it connected}, {\it totally disconnected} and {\it partially connected} paths. 
{\it Connected} paths are those where a state, $\ket{x_{a,b}}_i$ on step $i$ is followed by $\ket{x_{b,c}}_{i+1}$ on step $i+1$ for all the steps $i$. 
That is the second arrow index of step $i$ has to match the first arrow index of step $i+1$. 
If this property is not satisfied for every step on the path then we have a {\it totally disconnected} path, and the path is {\it partially connected} 
if this property is only satisfied for some of the steps. Examples of these three kinds of paths are shown in Fig.~\ref{dpaths}. 

\begin{figure}[h!]
\captionsetup{width=0.8\textwidth}
\begin{center}
		\includegraphics[scale=0.8]{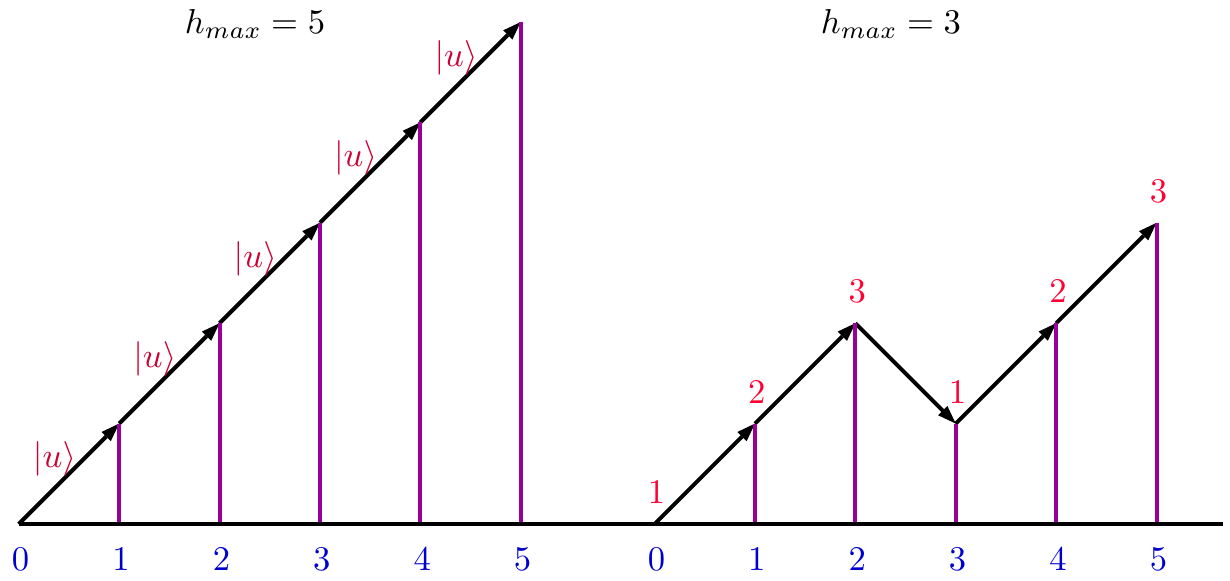} 
	\caption{\small Maximum heights reached in the DW and the modified DW for a 5-step walk.  
	 }
\label{heights_fredkin}
\end{center}
\end{figure}

\paragraph{Maximum heights} - Finally the maximum height we can reach in the modified DWs is no longer $n$ as in the DWs of $2n$ steps but 
\begin{equation}
h_{max} = \left[\frac{n-2}{3}\right] + 2,
\end{equation}
with $\left[k\right]$ being the greatest integer not exceeding $k$, (see Fig. \ref{heights_fredkin}). 

\subsection{The Hamiltonian $H_F$} \label{3}

\begin{figure}[h!]
\captionsetup{width=0.8\textwidth}
\begin{center}
		\includegraphics[scale=0.8]{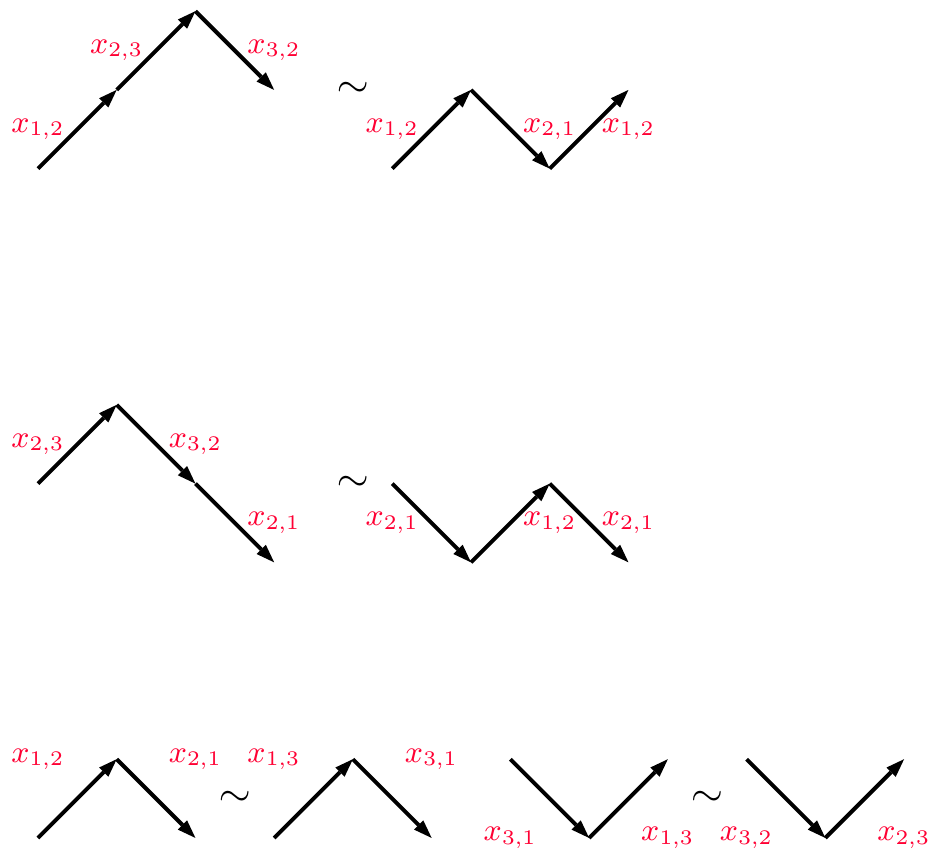} 
	\caption{\small The local equivalence moves for the modified Fredkin chain.}
\label{leSD}
\end{center}
\end{figure}

The different paths satisfying the conditions of the modified DWs can be mapped to each other using local equivalence moves illustrated in Fig. \ref{leSD}. 
We build a local, frustration-free Hamiltonian by projecting out these local moves, 
\begin{eqnarray}
\hspace{-7mm} U_{j, j+1, j+2} & = & \Pi^{\frac{1}{\sqrt{2}}\left[\ket{\left(x_{1,2}\right)_j, \left(x_{2,3}\right)_{j+1}, \left(x_{3,2}\right)_{j+2} } - \ket{\left(x_{1,2}\right)_j, \left(x_{2,1}\right)_{j+1}, \left(x_{1,2}\right)_{j+2}}\right]} \nn \\
\hspace{-7mm} D_{j, j+1, j+2} & = & \Pi^{\frac{1}{\sqrt{2}}\left[\ket{\left(x_{2,3}\right)_j, \left(x_{3,2}\right)_{j+1}, \left(x_{2,1}\right)_{j+2} } - \ket{\left(x_{2,1}\right)_j, \left(x_{1,2}\right)_{j+1}, \left(x_{2,1}\right)_{j+2}}\right]} \nn \\
\hspace{-7mm}W_{j, j+1} & = & \Pi^{\frac{1}{\sqrt{2}}\left[\ket{\left(x_{1,2}\right)_j, \left(x_{2,1}\right)_{j+1}} - \ket{\left(x_{1,3}\right)_j, \left(x_{3,1}\right)_{j+1}}\right]} \nn \\
& & + \lambda_1
\Pi^{\frac{1}{\sqrt{2}}\left[\ket{\left(x_{3,1}\right)_j, \left(x_{1,3}\right)_{j+1}} - \ket{\left(x_{3,2}\right)_j, \left(x_{2,3}\right)_{j+1}}\right]},
\end{eqnarray}
respectively with $\Pi^{\ket{\psi}}$ denoting a projector to the normalized state $\ket{\psi}$ and $\lambda_1 (\geq 0)$ a tunable parameter. 
Here $j\in\{1,\cdots, n-2\}$ for $U$ and $D$, and $j\in\{1,\cdots, n-1\}$ for $W$. Together they make up $H_{bulk, \,connected}$.

The boundary terms are given by  
\begin{eqnarray}
H_{left}  & = & \Pi^{\ket{\left(x_{2,1}\right)_1}} + \Pi^{\ket{\left(x_{3,1}\right)_1}} + \Pi^{ \ket{\left(x_{3,2}\right)_1}} +\Pi^{\ket{\left(x_{1,3}\right)_1, \left(x_{3,2}\right)_2, \left(x_{2,1}\right)_3}}, \\
H_{right}& = & \Pi^{\ket{\left(x_{1,2}\right)_n}}  + \Pi^{\ket{\left(x_{1,3}\right)_n}} + \Pi^{\ket{\left(x_{2,3}\right)_n}} +\Pi^{\ket{\left(x_{1,2}\right)_{n-2}, \left(x_{2,3}\right)_{n-1}, \left(x_{3,1}\right)_n}}, 
\end{eqnarray}
where the first three terms prevent the walks from moving below the $x$-axis at the origin and upward at $(n,0)$, and the last terms are new and have no analog in the original Fredkin model. 
They are added to suppress additional ground states not corresponding to the modified DWs~\footnote{
Without the last terms of $H_{left}$ and $H_{right}$, we can see that there are additional ground states except for the case starting and ending with the arrow index 2. 
For examples, $x_{1,3}x_{3,2}x_{2,1}x_{1,2}x_{2,3}x_{3,1}$ for length 6 and $x_{1,3}x_{3,2}x_{2,1}x_{1,3}x_{3,2}$ for length 5. 
As another choice of the boundary terms, we can restrict ground states to the case starting and ending at the arrow index 2 by 
\bea
H_{left}  & = & \Pi^{\ket{\left(x_{2,1}\right)_1}} + \Pi^{\ket{\left(x_{3,1}\right)_1}} + \Pi^{ \ket{\left(x_{3,2}\right)_1}} +\sum_{a=2}^3\Pi^{\ket{\left(x_{1,a}\right)_1}}, \\
H_{right}& = & \Pi^{\ket{\left(x_{1,2}\right)_n}}  + \Pi^{\ket{\left(x_{1,3}\right)_n}} + \Pi^{\ket{\left(x_{2,3}\right)_n}} +\sum_{a=2}^3\Pi^{\ket{\left(x_{a,1}\right)_n}}. 
\eea
}. 

We include a ``balancing'' term given by
\begin{equation}
B_{j, j+1} = \Pi^{\ket{\left(x_{1,3}\right)_j, \left(x_{3,2}\right)_{j+1}}} + \Pi^{\ket{\left(x_{2,3}\right)_j, \left(x_{3,1}\right)_{j+1}}}.
\end{equation} 
This term implies that if we go up with $\ket{\left(x_{1,3}\right)}$ or $\ket{\left(x_{2,3}\right)}$ we have to come down with $\ket{\left(x_{3,1}\right)}$ or $\ket{\left(x_{3,2}\right)}$, 
and is crucial for the phase transitions in this system. 

Finally we include the term
\be
H_{bulk, \,disconnected} = \sum_{j=1}^{n-1}\sum_{a,b,c, d = 1; a\neq b,\, b\neq c,\, c\neq d}^3~ \Pi^{\ket{\left(x_{a,b}\right)_j, \left(x_{c,d}\right)_{j+1}}}
\label{H_bulk_disconnected}
\ee
which keeps the disconnected paths out of the ground state sector.

With this the total Hamiltonian is
\begin{equation} \label{hs31}
H_F = H_{left} + H_{bulk, \,connected} +  H_{right} + \lambda_2\sum_{j=1}^{n-1}~B_{j, j+1} + H_{bulk, \,disconnected} .
\end{equation}
with $\lambda_2 (\geq 0)$ another tunable parameter. 
Notice that each of the terms in $H_{bulk, \,disconnected}$ commutes with the rest of $H_F$, block diagonalizing the Hamiltonian into inequivalent sectors distinguished by the number of disconnections.  

We can see that the Hamiltonian is frustration-free only when $\lambda_1$ or $\lambda_2$ vanishes. In what follows, let us consider these cases.

\section{Ground states } \label{sec:gs}

We discuss the structure of the ground states for the three cases 
\begin{itemize}
\item
$\lambda_1>0$, $\lambda_2=0$ 
\item
$\lambda_1=\lambda_2=0$
\item
$\lambda_1=0$, $\lambda_2> 0$ 
\end{itemize}
separately. 

\subsection{For $\lambda_1>0$, $\lambda_2=0$ :} \label{l0}

The arrow indices split the paths according to different equivalence classes that cannot be mapped into each other by local equivalence moves of Fig.~\ref{leSD}. 
That is if we start with the arrow index 1, that is with vectors $\ket{\left(x_{1,2}\right)}$ or $\ket{\left(x_{1,3}\right)}$ on the first step, 
we can end with either 1 or 2 as the arrow index on the last step. Thus we get two equivalences classes denoted by $\{11\}$ and $\{12\}$. 
In a similar manner we obtain the equivalence classes $\{22\}$, $\{21\}$, making the GSD 4, independent of the size of the chain. 
It is worth noting that this degeneracy does not arise due to the geometry or topology of the lattice but rather the additional degrees of freedom. 

In particular, there is no symmetry transformation mapping one equivalence class to another 
(except reversing the paths exchanging $\{12\}$ and $\{21\}$), which will be evident by seeing 
that the number of the paths in each equivalence class is different as (\ref{Nij_asym}). 
The absence of the symmetry comes from the constraint that the DWs are forbidden to enter the $y<0$ region. 

This GSD is stable to local perturbations in the bulk of the chain that preserve the local equivalence moves, 
but are sensitive to the boundary perturbations that can lift some of the states out of the ground state sector. 
For example, a local perturbation at a boundary by 
$\Pi^{\ket{\left(x_{1,2}\right)_1}} + \Pi^{\ket{\left(x_{1,3}\right)_1}}$ lifts $\{11\}$ and $\{12\}$ making the GSD 2. 

To understand the ground states in the different equivalence classes, we need to count the number of paths that satisfy
 the condition of the modified DWs which is the normalization of these states. 

 \subsubsection*{Normalization of the ground states for $\lambda_1>0$, $\lambda_2=0$}
 
\begin{figure}[h]
\captionsetup{width=0.8\textwidth}
\centering
\includegraphics[scale = 0.8]{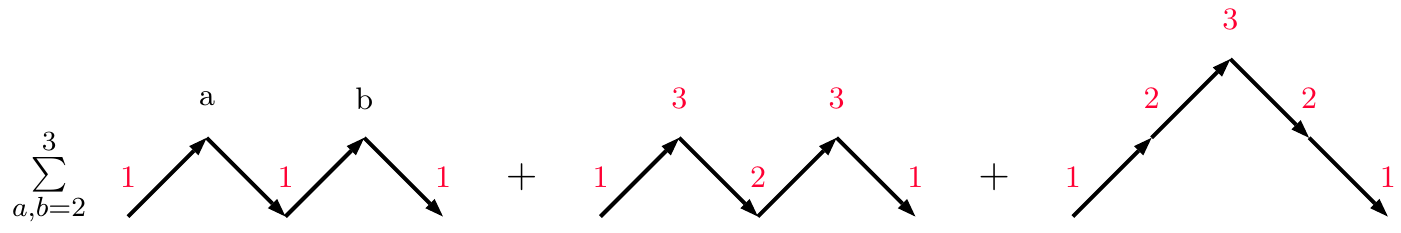}
\caption{\small Graphical expressions for the six terms of $P_{4,\, 1\to 1}$.} 
\label{fig:P4}
\end{figure}

 $P_{n,\,a\to b}$ denotes the formal sum of all possible connected paths on $n$ steps starting (ending) at the arrow index $a$ ($b$). 
For example, $P_{4,\, 1\to 1}$ is given as 
\begin{eqnarray} P_{4,\, 1\to 1} & = &  x_{1,2}x_{2,1}x_{1,2}x_{2,1} + x_{1,3}x_{3,1}x_{1,2}x_{2,1} + x_{1,2}x_{2,1}x_{1,3}x_{3,1} \nn \\ 
& + & x_{1,3}x_{3,1}x_{1,3}x_{3,1} + x_{1,3}x_{3,2}x_{2,3}x_{3,1} + x_{1,2}x_{2,3}x_{3,2}x_{2,1},\end{eqnarray}
shown in Fig.\ref{fig:P4}.

\noindent
\paragraph{Definition:}  
{\bf Let $N_{n,\,a\to b}$ be the number of walks included in $P_{n,\, a\to b}$, 
which is obtained by setting all the $x_{a,b}$ in $P_{n,\, a\to b}$ to 1. }
For instance, $N_{4,\, 1\to 1}=6$. 

We can see that 
$N_{n,\, a\to b}=N_{n,\, b\to a}$
by considering the reversed path starting from $(n,0)$ and ending at $(0,0)$. Also, $N_{n,\, 3\to a}=N_{n,\,a\to 3}=0$ with $a=1,2,3$. 
We use recursion relations to compute $N_{n, \, a\to b}$. 

\subsubsection*{Recursions for paths ending at height zero}
\begin{figure}[h]
\captionsetup{width=0.8\textwidth}
\centering
\includegraphics[scale = 0.8]{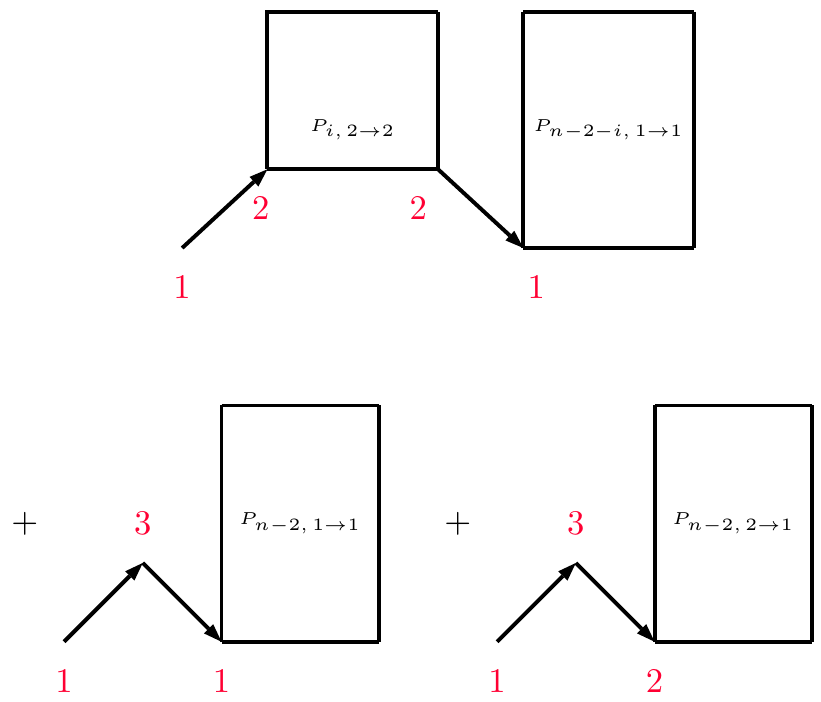}
\caption{\small The recursion in (\ref{P11_rec}) illustrated. } 
\label{recursion11_fredkin}
\end{figure}

By looking at the first step of the walks, we can write down the following recursions (for example see Fig.~\ref{recursion11_fredkin}):
\bea
P_{n,\, 1\to 1} & = & x_{1,2}\sum_{i=0}^{n-2}P_{i,\,2 \to 2}\, x_{2,1}\,P_{n-2-i,\,1\to 1} \nn \\
& &   +x_{1,3}\, x_{3,1}\,P_{n-2,\,1\to 1} +x_{1,3}\, x_{3,2}\,P_{n-2,\,2\to 1} , 
\label{P11_rec}
\\
P_{n,\, 2\to 2} & = & x_{2,3}\, x_{3,2}\,P_{n-2,\,2\to 2} +x_{2,3}\, x_{3,1}\,P_{n-2,\,1\to 2}, 
\label{P22_rec}
\\
P_{n,\, 2\to 1} & = & x_{2,3}\, x_{3,2}\,P_{n-2,\,2\to 1} +x_{2,3}\, x_{3,1}\,P_{n-2,\,1\to 1}. 
\label{P21_rec}
\eea

These lead to recursions for $N_{n,\, a\to b}$ as~\footnote{These are valid for $n\geq 1$. The terms of $\sum_{i=0}^{n-2}$ are regarded as null for $n=1$. } 
\bea
N_{n,\, 1\to 1} & = & \sum_{i=0}^{n-2}N_{i,\,2 \to 2}\, N_{n-2-i,\,1\to 1}  +N_{n-2,\,1\to 1} +N_{n-2,\,2\to 1} , 
\label{N11_rec}
\\
N_{n,\, 2\to 2} & = & N_{n-2,\,2\to 2}  +N_{n-2,\,2\to 1}, 
\label{N22_rec}
\\
N_{n,\, 2\to 1} & = & N_{n-2,\,2\to 1} +N_{n-2,\,1\to 1}, 
\label{N21_rec}
\eea
where we use the invariance under the reversal property of the paths. By introducing the generating functions~\footnote{
The initial conditions $N_{0,\, a\to b}= \delta_{a, b}$ are determined by (\ref{N11_rec})-(\ref{N21_rec}) at $n=2$.
} 
\be
N_{a\to b}(x)\equiv \sum_{n=0}^\infty N_{n,\, a\to b}\, x^n \qquad \mbox{with} \qquad N_{0,\, a\to b}= \delta_{a, b},
\label{Nijg}
\ee
(\ref{N11_rec})-(\ref{N21_rec}) are recast as 
\bea
N_{1\to 1}(x) -1 & = & x^2N_{2\to 2}(x)N_{1\to 1}(x) +x^2N_{1\to 1}(x) +x^2 N_{2\to 1}(x) , 
\label{N11g_rec}
\\
N_{2\to 2}(x) -1 & = & x^2N_{2\to 2}(x) +x^2 N_{2\to 1}(x) , 
\label{N22g_rec}
\\
N_{2\to 1}(x) & = & x^2N_{2\to 1}(x) +x^2 N_{1\to 1}(x) ,
\label{N21g_rec}
\eea
which are solved as 
\bea
N_{1\to 1}(x) & = & \frac{1-x^2}{2x^3X}\left[1-\sqrt{1-4X^2}\right], 
\label{N11g_sol}
\\
N_{2\to 2}(x) & = & \frac{1}{1-x^2}\left[1+ \frac{x}{2X}\left(1-\sqrt{1-4X^2}\right)\right], 
\label{N22g_sol}
\\
N_{2\to 1}(x) & = & \frac{1}{2xX}\left[1-\sqrt{1-4X^2}\right]
\label{N21g_sol}
\eea
with 
\be 
X\equiv \frac{x^3}{1-3x^2}. 
\label{X}
\ee

Among the singularities of (\ref{N11g_sol}), (\ref{N22g_sol}) and (\ref{N21g_sol}), 
the nearest from the origin is $x=\pm 1/2$. Around these points, they behave as 
\bea
N_{1\to 1}(x) & = & 6-18\sqrt{2}\sqrt{1\mp 2x} + O(1\mp 2x), \nn \\
N_{2\to 2}(x) & = & 2-2\sqrt{2}\sqrt{1\mp 2x} + O(1\mp 2x), \nn \\
N_{2\to 1}(x) & = & 2-6\sqrt{2}\sqrt{1\mp 2x}+ O(1\mp 2x). 
\label{Nijg_sing}
\eea
Since (\ref{N11g_sol})-(\ref{N21g_sol}) are even functions of $x$, we should equally take into account both contributions from $x=1/2$ and $x=-1/2$, 
in order to obtain the large order behavior of coefficients. 
Looking at the large order behavior in the expansion of $\sqrt{1\mp 2x}$ around $x=0$, 
we can read off the large order behavior of the coefficients:
\bea
& & N_{n,\, 1\to 1}  \sim  \frac{1+(-1)^n}{2}\,\frac{18\sqrt{2}}{\sqrt{\pi}}\frac{2^n}{n^{3/2}}, \qquad 
N_{n, \,2\to 2}   \sim  \frac{1+(-1)^n}{2}\,\frac{2\sqrt{2}}{\sqrt{\pi}}\frac{2^n}{n^{3/2}}, \nn \\
& & N_{n, \,2\to 1}   \sim  \frac{1+(-1)^n}{2}\,\frac{6\sqrt{2}}{\sqrt{\pi}}\frac{2^n}{n^{3/2}} 
\label{Nij_asym}
\eea
as $n\to \infty$. 

\subsubsection*{Recursions for paths ending at nonzero height}
\begin{figure}[h]
\captionsetup{width=0.8\textwidth}
\centering
\includegraphics[scale = 0.8]{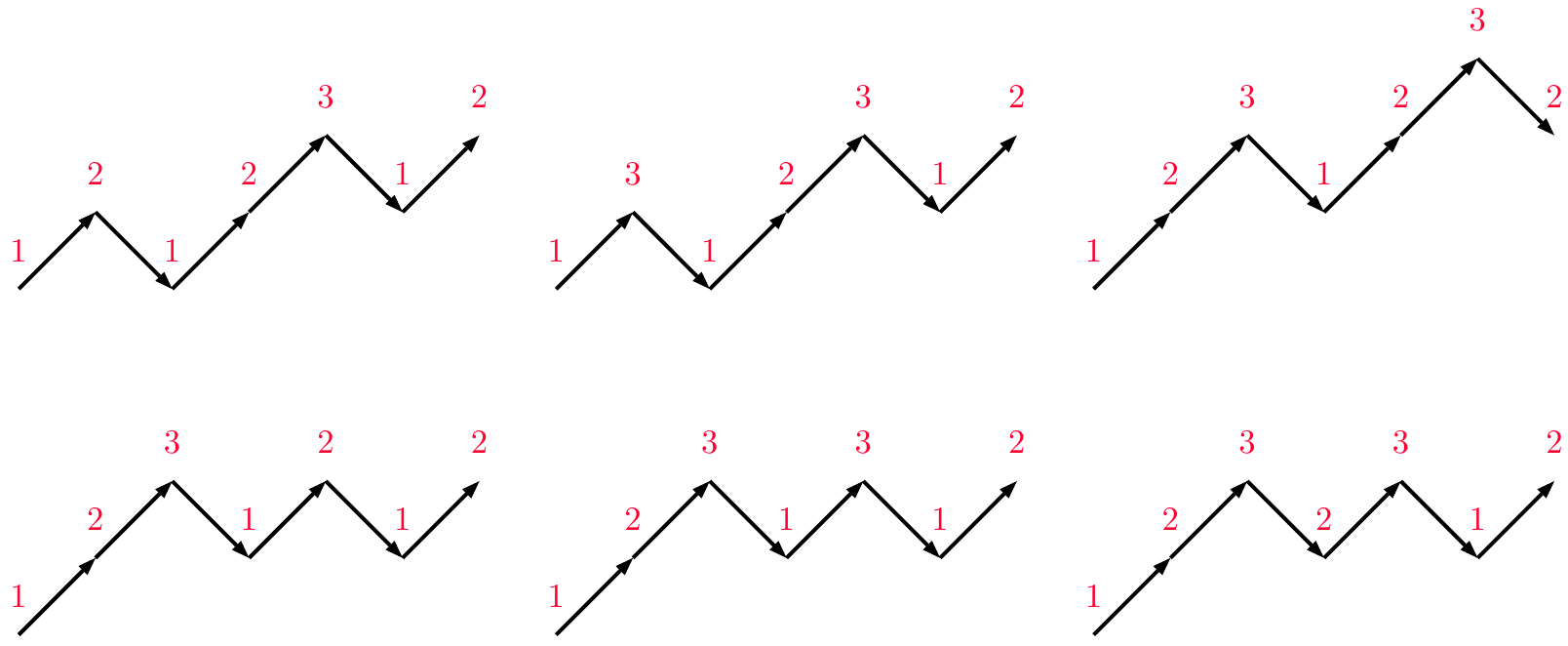}
\caption{\small Graphical expressions for the six terms of $P^{(2)}_{6,\, 1\to 2}$. } 
\label{fig:P(2)6}
\end{figure}

For later convenience, we also consider $n$-step walks obeying similar rules but starting at $(0,0)$ with the arrow index $a$ and ending at $(n,h)$ with the index $b$. 
$h$ is a positive integer, and the paths never pass below the $x$-axis. 
$P^{(h)}_{n,\,a\to b}$ denotes the sum of such walks, and $N^{(h)}_{n,\, a\to b}$ counts the number of the walks in $P^{(h)}_{n,\, a\to b}$. 
For example, 
\bea
P^{(2)}_{6,\,1\to 2} & = & \left(x_{1,2}x_{2,1} + x_{1,3}x_{3,1}\right) x_{1,2} x_{2,3}x_{3,1}x_{1,2} + x_{1,2}x_{2,3}x_{3,1}x_{1,2}x_{2,3}x_{3,2} \nn \\
& & + x_{1,2} x_{2,3}x_{3,1} \left(x_{1,2} x_{2,1} + x_{1,3}x_{3,1}\right) x_{1,2}+ x_{1,2}x_{2,3}x_{3,2} x_{2,3}x_{3,1}x_{1,2}, 
\label{P(2)6}\\
N^{(2)}_{6,\, 1\to 2} & = & 6.
\eea 
The six paths in the r.h.s. of (\ref{P(2)6}) are depicted in Fig.~\ref{fig:P(2)6}. 
It is easy to see 
\be
N^{(h)}_{n,\, 3\to b}=0 \qquad \mbox{for} \quad b=1,2,3, \quad \mbox{and} \quad h\geq 1. 
\ee
Namely, there exists no path starting with the arrow index 3 for any positive height. 

\begin{figure}[h]
\captionsetup{width=0.8\textwidth}
\centering
\includegraphics[scale = 0.8]{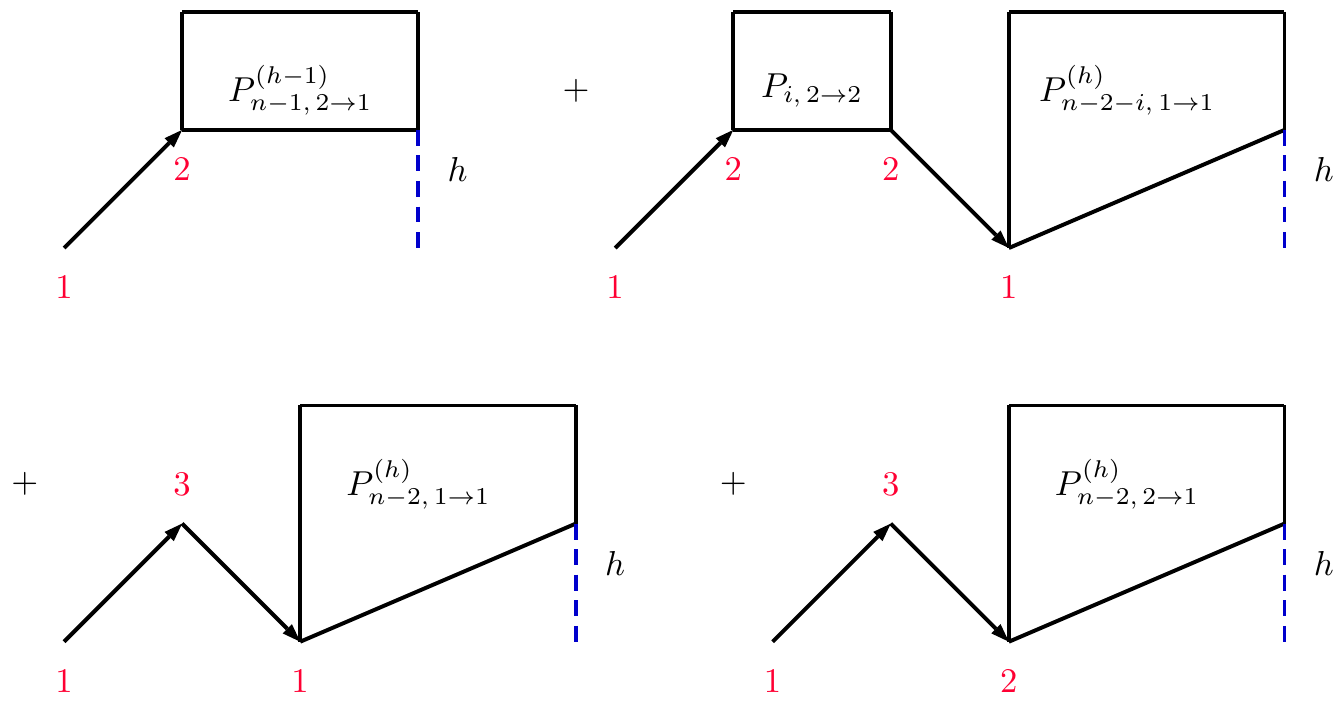}
\caption{\small The recursion in (\ref{Nh1j_rec}) illustrated for the $b=1$ case. } 
\label{recursion11h_fredkin}
\end{figure}

We obtain recursion relations for these walks similar to the case of zero height. This is illustrated in Fig. \ref{recursion11h_fredkin} for the $N^{(h)}_{n,\, 1\to 1}$ case.
The result is~\footnote{
Here, $N^{(0)}_{n,\, a\to b}$ is regarded as $N_{n,\, a\to b}$.
}   
\bea
N^{(h)}_{n,\, 1\to b} & = & N^{(h-1)}_{n-1,\, 2\to b} + \delta_{b, 3}\delta_{h,1}\delta_{n,1} +\sum_{i=0}^{n-2}N_{i,\, 2\to 2}\, N^{(h)}_{n-2-i,\,1\to b} \nn \\
& & +N^{(h)}_{n-2,\,1\to b} + N^{(h)}_{n-2,\,2\to b} , 
\label{Nh1j_rec}
\\
N^{(h)}_{n,\, 2\to b} & = & \delta_{b, 3}\delta_{h,1}\delta_{n,1}  +N^{(h)}_{n-2,\,1\to b}  + N^{(h)}_{n-2,\,2\to b} , 
\label{Nh2j_rec}
\eea
for $b=1,2,3$, $n\geq 1$ and $h\geq 1$. 
In terms of the generating functions 
\be
N^{(h)}_{a\to b}(x)\equiv \sum_{n=0}^\infty N^{(h)}_{n,\, a\to b}\, x^n \qquad \mbox{with} \qquad N^{(h)}_{0,\,a\to b}=0
\label{Nhijg}
\ee
together with (\ref{Nijg}), we find that the pair of the equations (\ref{Nh1j_rec}) and (\ref{Nh2j_rec}) is closed for each $b$. 
Noting the relation from the zero height case 
\be
N^{(0)}_{2 \to b}(x)=\frac{x^2}{1-x^2}N^{(0)}_{1\to b}(x) + \delta_{b,2}\frac{1}{1-x^2},
\ee
we eventually obtain 
\bea
N^{(h)}_{2\to 1}(x) & = & \frac{1}{x}\left(\frac{x^3}{1-x^2} \,N_{1\to 1}(x)\right)^{h+1}, 
\label{Nh21g_sol}
\\
N^{(h)}_{1\to 1}(x) & = &  \frac{1-x^2}{x^3}\left(\frac{x^3}{1-x^2} \,N_{1\to 1}(x)\right)^{h+1}, 
\label{Nh11g_sol}
\\
N^{(h)}_{2\to 2}(x) & = & \frac{1}{1-x^2}\left(\frac{x^3}{1-x^2} \,N_{1\to 1}(x)\right)^{h}\left\{\frac{x^4}{1-x^2}N_{1\to 1}(x) +1\right\}, 
\label{Nh22g_sol}
\\
N^{(h)}_{1\to 2}(x) & = & \frac{1}{x^2}\left(\frac{x^3}{1-x^2} \,N_{1\to 1}(x)\right)^{h}\left\{\frac{x^4}{1-x^2}N_{1\to 1}(x) +1\right\}, 
\label{Nh12g_sol}
\\
N^{(h)}_{2\to 3}(x) & = & \frac{x}{1-x^2}\left(\frac{x^3}{1-x^2} \,N_{1\to 1}(x)\right)^{h-1}\left\{\frac{x^2}{1-x^2}N_{1\to 1}(x)+1\right\},
\label{Nh23g_sol}
\\
N^{(h)}_{1\to 3}(x) & = & -\delta_{h,1}\frac{1}{x}+\frac{1}{x}\left(\frac{x^3}{1-x^2} \,N_{1\to 1}(x)\right)^{h-1}\left\{\frac{x^2}{1-x^2}N_{1\to 1}(x)+1\right\}. 
\label{Nh13g_sol}
\eea
Plugging (\ref{Nijg_sing}) into these, we can read off the large order behavior of $N^{(h)}_{n,\, a\to b}$, which is useful for a fixed $h$ as $n\to \infty$ but not for cases where
both $n$ and $h$ are growing. 
In order to find useful expressions even for the latter cases, we use the identity:
\be
X^h\left\{\frac{1}{2X^2}\left(1-\sqrt{1-4X^2}\right)\right\}^{h+1}=\sum_{n=0}^\infty N_n^{(h)}X^n 
\label{Id}
\ee
(see eq.(\ref{X})) with
\be
N^{(h)}_n=\frac{1+(-1)^{n+h}}{2}\,\frac{h+1}{\frac{n+h}{2}+1} \binomi{n}{\frac{n+h}{2}}.
\label{Nhn}
\ee
This is derived in appendix A in \cite{FSPP} by computing the number of the DWs in two different ways.

\subsubsection*{$\boldsymbol{N^{(h)}_{p,\, 2\to 1}}$ :} 
First, let us obtain the large order behavior of $N^{(h)}_{p,\, 2\to 1}$ as $p\to \infty$. 
Applying (\ref{Id}) to (\ref{Nh21g_sol}) with (\ref{N11g_sol}) leads to
\be
N^{(h)}_{2\to 1}(x)=\frac{1}{x}\sum_{n=0}^\infty N^{(h)}_nX^{n+1}. 
\ee
From (\ref{X}), $X^{n+1}$ has the expansion 
\be
X^{n+1}= \sum_{k=0}^\infty \binomi{n+k}{k} 3^k \,x^{3n+2k+3}.
\ee
Plugging these two, we have 
\be
N^{(h)}_{p,\,2\to 1}=\sum_{n\geq 0}^*N^{(h)}_n\binomi{\frac{p-n}{2}-1}{n} 3^{\frac{p-3n}{2}-1} ,
\label{Nh21_sol}
\ee
where the asterisk (*) on the summation means $n$ running under the condition for $p-3n$ to be even and 
no less than 2. 
It is found that the summand of (\ref{Nh21_sol}) has a saddle point (a stable point with respect to the deviation $n\to n+2$) at
\be
n\sim\frac{1}{9}p
\label{saddle}
\ee
for $p$ large. 

By using Stirling's formula ($n!\simeq \sqrt{2\pi}\,n^{n+\frac12}e^{-n}$), 
the asymptotic form of $N^{(h)}_n$ becomes 
\bea
N^{(h)}_n & \simeq & \frac{1+(-1)^{n+h}}{2}\,(h+1)\frac{2^{3/2}}{\sqrt{\pi}}\frac{2^n}{n^{3/2}}\nn \\
& & \times\exp\left[1-\frac{n+h+3}{2}\ln\left(1+\frac{h+2}{n}\right)-\frac{n-h+1}{2}\ln\left(1-\frac{h}{n}\right)\right] \nn \\
& & \times \left[1+O\left(n^{-1}\right)\right]. 
\eea
The power of the exponential is expanded in large $n$ as 
\be
-\frac{(h+1)^2}{2n}-\frac{3}{2n}+\frac{(h+1)^2}{n^2} + \frac{2}{3n^2} +O\left(\frac{h^4}{n^3}\right), 
\ee
in which the first term provides the Gaussian factor rapidly decaying for $h> \sqrt{n}$. 
Bringing down the other terms from the exponential, we obtain  
\be
N^{(h)}_n\simeq \frac{1+(-1)^{n+h}}{2}\,(h+1)\frac{2^{3/2}}{\sqrt{\pi}}\frac{2^n}{n^{3/2}}\, e^{-\frac{1}{2n}(h+1)^2} 
\times \left[1+\frac{(h+1)^2}{n^2} + O\left(\frac{1}{n},\,\frac{h^4}{n^3}\right) \right]. 
\label{Nhn_asym0}
\ee
Note that due to the Gaussian factor the order of $h$ is effectively at most $O\left(\sqrt{n}\right)$. So, we can regard 
the terms of $\frac{(h+1)^2}{n^2}$ and $\frac{h^4}{n^3}$ as $O\left(n^{-1}\right)$ quantities. 
(\ref{Nhn_asym0}) can be written as 
\bea
N^{(h)}_n & \simeq & \frac{1+(-1)^{n+h}}{2}\,(h+1)e^{-\frac{1}{2n}(h+1)^2} \times \left\{\begin{matrix} N^{(0)}_n & \text{($h$: even)} \\ \frac12 N^{(1)}_n & \text{($h$: odd)} \end{matrix}\right\}\nn \\
 & & \times \left[1+O\left(n^{-1}\right)\right]
\label{Nhn_asym}
\eea
with $N^{(0)}_n\sim \frac12N^{(1)}_n\sim\frac{2^{3/2}}{\sqrt{\pi}}\frac{2^n}{n^{3/2}}\times \left[1+O\left(n^{-1}\right)\right]$. 
We plug (\ref{Nhn_asym}) to (\ref{Nh21_sol}) and evaluate the sum over $n$ in the saddle point method around (\ref{saddle}). 
For the case of $h$ even, by putting $h=2k$, $n=2m$ and $p=2q$, (\ref{Nh21_sol}) is expressed as 
\be
N^{(2k)}_{2q,\,2\to 1}\simeq (2k+1)\sum_{m=0}^{\left[\frac{q-1}{3}\right]}e^{-\frac{(2k+1)^2}{4m}} N^{(0)}_{2m}\binomi{q-m-1}{2m} 3^{q-3m-1}.
\ee
After expanding the summand around the saddle point 
$m=\frac19q+x$
with $x$ denoting the fluctuation, we have 
\bea
N^{(2k)}_{2q,\,2\to 1} & \simeq & (2k+1)e^{-\frac{9}{4q}(2k+1)^2}\,\frac{3^{7/2}}{4\pi}\frac{2^{2q}}{q^2}\int^\infty_{-\infty}dx \,e^{-\frac{9\cdot 27}{16q}x^2} \nn \\
& & \times \left[1-\frac{27}{16q}x+O\left(\frac{x^2}{q^2},\, \frac{x^3}{q^2}\right)\right]\nn \\
& & \times \exp\left[\frac{9^2}{4q^2}(2k+1)^2x -\frac{9^3}{4q^3}(2k+1)^2x^2+\cdots\right], 
\eea
where the sum is converted to the integral. The second line comes from the fluctuation in the factor $N^{(0)}_{2m}\binomi{q-m-1}{2m} 3^{q-3m-1}$, while the third line from the fluctuation in $e^{-\frac{(2k+1)^2}{4m}}$. 
The integral is evaluated by expanding the exponential in the third line. The linear terms in $x$ appear because $\frac19q$ is an approximate saddle point. Although they are at most of the order $O(q^{-1/2})$, they do not contribute 
to the integral due to the parity. It is easy to see that higher orders in $x$ yield contributions at most to $O(q^{-1})$, and the final result takes the form: 
\be
N^{(2k)}_{2q,\,2\to 1}\simeq (2k+1) e^{-\frac{9}{4q}(2k+1)^2}\,\frac{3\cdot 2^{3/2}}{\sqrt{\pi}}\frac{2^{2q}}{(2q)^{3/2}}\times \left[1+O(q^{-1})\right].
\ee
We can evaluate similarly for the case of $h$ odd, and summarize the results for both of the cases as  
\be
N^{(h)}_{p,\, 2\to 1} \sim  \frac{1+(-1)^{p+h}}{2}\, (h+1)e^{-\frac{9}{2p}(h+1)^2} \, \frac{6\sqrt{2}}{\sqrt{\pi}}\frac{2^p}{p^{3/2}} \times \left[1+O\left(p^{-1}\right)\right].
\label{Nh21_asym}
\ee

\subsubsection*{Other coefficients:}
Once we know (\ref{Nh21_asym}), it is straightforward to obtain the large order behavior for the other coefficients from (\ref{Nh21g_sol})-(\ref{Nh13g_sol}). 
For instance, we find $N^{(h)}_{p,\,1\to 1}= N^{(h)}_{p+2,\,2\to 1}-N^{(h)}_{p,\,2\to 1}$ from (\ref{Nh11g_sol}). 
Eventually, we have the expressions (up to multiplicative factors of $\left[1+O\left(p^{-1}\right)\right]$):
\bea
N^{(h)}_{p,\,1\to 1} 
& \sim & \frac{1+(-1)^{p+h}}{2}\, (h+1) e^{-\frac{9}{2p}(h+1)^2} \, \frac{18\sqrt{2}}{\sqrt{\pi}}\frac{2^p}{p^{3/2}},
\label{Nh11_asym}
\\
N^{(h)}_{p,\,1\to 2}  
& \sim & \frac{1+(-1)^{p+h}}{2}\, \left[2h e^{-\frac{9}{2p}h^2}+(h+1) e^{-\frac{9}{2p}(h+1)^2} \right]\frac{6\sqrt{2}}{\sqrt{\pi}}\frac{2^p}{p^{3/2}},
\label{Nh12_asym}
\\
N^{(h)}_{p,\, 2\to 2}&\sim & \frac{1+(-1)^{p+h}}{2}\, \left[2h e^{-\frac{9}{2p}h^2}+(h+1) e^{-\frac{9}{2p}(h+1)^2} \right]\frac{2\sqrt{2}}{\sqrt{\pi}}\frac{2^p}{p^{3/2}},
\label{Nh22_asym}
\\
N^{(h)}_{p,\,2\to 3}  
& \sim & \frac{1+(-1)^{p+h}}{2}\, \left[2h e^{-\frac{9}{2p}h^2}-(h-1) e^{-\frac{9}{2p}(h-1)^2} \right]\frac{2\sqrt{2}}{\sqrt{\pi}}\frac{2^p}{p^{3/2}},
\label{Nh23_asym}
\\
N^{(h)}_{p,\,1\to 3}  
& \sim &\frac{1+(-1)^{p+h}}{2}\,  \left[2h e^{-\frac{9}{2p}h^2}-(h-1) e^{-\frac{9}{2p}(h-1)^2} \right]\frac{6\sqrt{2}}{\sqrt{\pi}}\frac{2^p}{p^{3/2}}
\label{Nh13_asym}
\eea
for $h\geq 1$. 
 
 As a consistency check, we can see that (\ref{Nh21_asym})-(\ref{Nh13_asym}) together with (\ref{Nij_asym}) satisfy the composition law~\footnote{
Note that the number of $p$-step paths from height $h$ with the arrow index $b$ to height 0 with the index $c$ is equal to $N^{(h)}_{p, \,c\to b}$. 
The sum over $h$ can be computed by converting it to the integral. 
}:
 \be
 \sum_{h=0}^\infty\sum_{b=1}^3 N^{(h)}_{p+r,\,a\to b} N^{(h)}_{p-r,\,c\to b} =N_{2p,\, a\to c} 
 \label{composition}
 \ee
for both of $p+r$ and $p-r$ being $O(p)$. This holds except for errors of $O\left(p^{-1}\right)$. 

\subsection{For $\lambda_1=\lambda_2=0$ : }

In this case, there is no equivalence move of $x_{3,1}\,x_{1,3} \sim x_{3,2}\, x_{2,3}$, which amounts to a degeneracy in each of the equivalence classes, $\{11\}$, $\{12\}$, $\{21\}$ and $\{22\}$. For example, 
for length-4 paths, each of the four sectors splits into two inequivalent paths: 
\bea
\{11\} &\to & \left(x_{1,2}x_{2,1}+x_{1,3}x_{3,1}\right)^2 + x_{1,2}x_{2,3}x_{3,2}x_{2,1}, \quad x_{1,3}x_{3,2}x_{2,3}x_{3,1}, 
\label{length4-11}\\
\{12\} & \to & x_{1,2}x_{2,1}x_{1,3}x_{3,2}+x_{1,3}x_{3,1}x_{1,3}x_{3,2}, \quad x_{1,3}x_{3,2}x_{2,3}x_{3,2}, \\
\{21\} & \to & x_{2,3}x_{3,1}x_{1,2}x_{2,1}+x_{2,3}x_{3,1}x_{1,3}x_{3,1}, \quad x_{2,3}x_{3,2}x_{2,3}x_{3,1},\\
\{22\} & \to & x_{2,3}x_{3,1}x_{1,3}x_{3,2},\quad \left(x_{2,3}x_{3,2}\right)^2, 
\eea
making the total GSD 8. 

For the $\{11\}$ sector, we can see that any length-$2n$ path obtained from $\left(x_{1,2}x_{2,1}\right)^n$ by successively applying the allowed equivalence moves 
cannot exceed height 2. In all of these paths, the arrow indices 1 and 2 appear only at height 0 and 1 respectively, and the index 3 at height 1 or 2. 
The first part of (\ref{length4-11}) gives the case of $n=2$. 

\begin{figure}[h]
\captionsetup{width=0.8\textwidth}
\centering
\includegraphics[scale = 0.8]{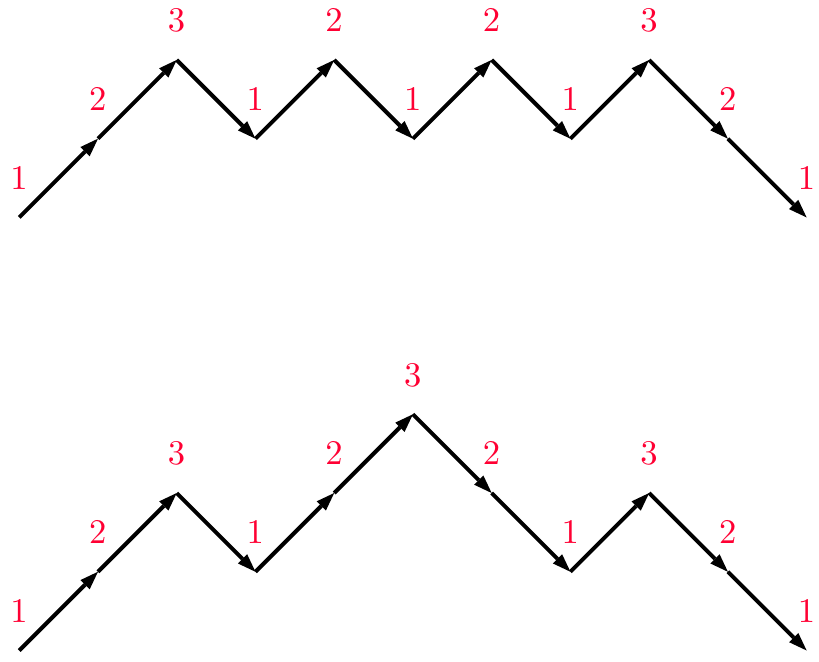}
\caption{\small The path (\ref{length10-1}) is connected to the path (\ref{length10-2}) of height 3 by the equivalence moves. } 
\label{path10} 
\end{figure}

To expose the degeneracy we start with paths inequivalent to $\left(x_{1,2}x_{2,1}\right)^n$ and obtain paths exceeding height 2. 
For example, consider a length-10 path: 
\be
x_{1,2}x_{2,3}x_{3,1}\left(x_{1,2}x_{2,1}\right)^2 x_{1,3}x_{3,2}x_{2,1} . 
\label{length10-1}
\ee
The equivalence moves provide the path that reaches height 3:
\be
x_{1,2}x_{2,3}x_{3,1}x_{1,2}x_{2,3}x_{3,2}x_{2,1}x_{1,3}x_{3,2}x_{2,1}
\label{length10-2}
\ee
as in Fig.~\ref{path10}. 
Note that 
$x_{1,2}x_{2,3}x_{3,1}$ and $x_{1,3}x_{3,2}x_{2,1} $ in the left and right edges are unchanged by the moves, and only the middle part is affected. 
What is relevant for the EE is the middle part. 
As a result of the equivalence moves, that part gives rise to the first of the two inequivalent paths in (\ref{length4-11}), and the problem reduces to the case of $\left(x_{1,2}x_{2,1}\right)^2$. 
This observation can be generalized to longer paths. 
In addition, if we start with the following path of length $2(n_1+n_2)+4$:
\be
\left(x_{1,2}x_{2,1}\right)^{n_1} x_{1,3}x_{3,2}x_{2,3}x_{3,1}\left(x_{1,2}x_{2,1}\right)^{n_2}, 
\ee
the middle part $x_{1,3}x_{3,2}x_{2,3}x_{3,1}$ is unchanged by the moves, and the two parts $\left(x_{1,2}x_{2,1}\right)^{n_1}$ and $\left(x_{1,2}x_{2,1}\right)^{n_2}$ 
are separately affected by the moves. This again reduces to the case $\left(x_{1,2}x_{2,1}\right)^{n}$. 

This is also valid for the other sectors. 
For sectors of $\{12\}$, $\{21\}$ and $\{22\}$, length-$2n$ paths obtained from
\be
 \left(x_{1,2}x_{2,1}\right)^{n-1}x_{1,3}x_{3,2}, \quad x_{2,3}x_{3,1} \left(x_{1,2}x_{2,1}\right)^{n-1}, \quad \mbox{and}\quad 
 x_{2,3}x_{3,1} \left(x_{1,2}x_{2,1}\right)^{n-2}x_{1,3}x_{3,2}
 \ee
are the most relevant, respectively. Since the edge part $x_{1,3}x_{3,2}$ or $x_{2,3}x_{3,1}$ does not change through the moves, 
the problem reduces to the case $\left(x_{1,2}x_{2,1}\right)^n$. 
Thus, although we have numerous GSD growing with the length in the case of $\lambda_1=\lambda_2=0$, we can conclude that the ground state corresponding to paths 
obtained from $\left(x_{1,2}x_{2,1}\right)^n$ through the equivalence moves is the most relevant for the computation of the EE. 
So we will only compute the EE of the state arising from acting on $\left(x_{1,2}x_{2,1}\right)^n$ with the equivalence moves. 

\subsubsection*{Normalization of the ground states for $\lambda_1=\lambda_2=0$}
Let $Q_{2n,\,1\to 1}$ be the formal sum of all possible length-$2n$ paths obtained from $\left(x_{1,2}x_{2,1}\right)^n$ by the equivalence moves. 
We can write down a recursion relation as
\bea
Q_{2n,\,1\to 1} & = & \left(x_{1,2}x_{2,1}+x_{1,3}x_{3,1}\right)Q_{2n-2,\,1\to 1} \nn \\
& & +\sum_{k=1}^{n-1}x_{1,2}\left(x_{2,3}x_{3,2}\right)^k\,x_{2,1}\,Q_{2n-2k-2,\,1\to 1}
\label{Qrecursion}
\eea    
with $Q_{0,\,1\to 1}=1$. 
For $M_{2n,\,1\to 1}$ denoting the number of paths in $Q_{2n,\,1\to 1}$, the corresponding recursion reads 
\be
M_{2n\,1\to 1}=2M_{2n-2,\,1\to 1} + \sum_{k=1}^{n-1}M_{2n-2k-2,\,1\to 1} 
\label{Mrecursion}
\ee
for $n\geq 1$. By introducing the generating function
\be
M_{1\to 1}(y)\equiv \sum_{n=0}^\infty M_{2n,\,1\to 1}\,y^n \qquad \mbox{with}\qquad  M_{0,\,1\to 1}=1,
\label{My}
\ee
(\ref{Mrecursion}) is recast as 
\be
M_{1\to 1}(y)-1=2yM_{1\to 1}(y)+\frac{y^2}{1-y}M_{1\to 1}(y).
\label{Myeq}
\ee
Here, we treated the double sum $\sum_{m=1}^\infty\sum_{\ell=0}^{m-1}M_{2\ell,\,1\to 1}\,y^m$ 
by changing the order of the sums as $\sum_{m=1}^\infty\sum_{\ell=0}^{m-1}\cdots=\sum_ {\ell=0}^\infty\sum_{m=\ell+1}^\infty\cdots$. 
The solution of (\ref{Myeq}) reads 
\be
M_{1\to 1}(y)= \frac{1-y}{1-3y+y^2}.
\label{Mysol} 
\ee 
The singularity of $M_{1\to 1}(y)$ nearest from the origin is 
\be
y_-=\frac{3-\sqrt{5}}{2}=\left(\frac{\sqrt{5}-1}{2}\right)^2, 
\ee
around which $M_{1\to 1}(y)$ behaves as 
\be
M_{1\to 1}(y)=\frac{\sqrt{5}-1}{2\sqrt{5}}\frac{1}{y_--y}+O\left((y_--y)^0\right). 
\ee
This determines the asymptotic behavior of the coefficient $M_{2n,\,1\to 1}$ as 
\be
M_{2n,\,1\to 1}\sim \frac{\sqrt{5}-1}{2\sqrt{5}}\frac{1}{y_-^{n+1}}=\frac{1}{\sqrt{5}}\left(\frac{\sqrt{5}+1}{2}\right)^{2n+1}
\label{M2nasymp}
\ee

\subsubsection*{Paths ending at nonzero height}
In order to compute the EE in section~\ref{sec:EE00}, we consider paths ending at nonzero height ($h=1,2$). 
For $h=1$, let $Q^{(1)}_{2n+1, \,1\to a}$ with $a=2,3$ be the sum of paths obtained from $\left(x_{1,2}x_{2,1}\right)^n\,x_{1,a}$ by the equivalence moves, 
and for $h=2$, $Q^{(2)}_{2n,\,1\to 3}$ denotes the sum of paths obtained from $\left(x_{1,2}x_{2,1}\right)^{n-1}\,x_{1,2}x_{2,3}$. 
 
By looking at final steps, we find the recursion relations as 
\bea
Q^{(1)}_{2n+1,\,1\to 2} & = & \sum_{k=0}^nQ_{2n-2k,\,1\to 1}\,x_{1,2}\left(x_{2,3}x_{3,2}\right)^k, 
\label{Q1recursion1}\\
Q^{(1)}_{2n+1,\,1\to 3} & = & Q_{2n,\, 1\to 1}\,x_{1,3},
\label{Q1recursion2}\\
Q^{(2)}_{2n,\,1\to 3} & = & Q^{(1)}_{2n-1,\,1\to 2}\,x_{2,3}.
\label{Q2recursion}
\eea
For the number of paths in $Q^{(1)}_{2n+1,\,1\to 2}$, denoted by $M^{(1)}_{2n+1,\,1\to 2}$, the corresponding equation can be solved in a similar manner to the zero height case. 
The result is
\be
M^{(1)}_{2n+1,\,1\to 2}\sim \frac{1}{\sqrt{5}}\left(\frac{\sqrt{5}+1}{2}\right)^{2n+2}.
\label{M112asymp}
\ee
Then, the numbers of paths in the other two, $M^{(1)}_{2n+1,\,1\to 3}$ and $M^{(2)}_{2n,\,1\to 3}$, are 
\bea
 & & M^{(1)}_{2n+1,\,1\to 3}=M_{2n,\,1\to 1}\sim \frac{1}{\sqrt{5}}\left(\frac{\sqrt{5}+1}{2}\right)^{2n+1}, 
 \label{M113asymp}
 \\
& & M^{(2)}_{2n,\,1\to 3} = M^{(1)}_{2n-1,\,1\to 2}\sim \frac{1}{\sqrt{5}}\left(\frac{\sqrt{5}+1}{2}\right)^{2n}.
\label{M213asymp}
\eea

As a consistency check, we can see that the results satisfy the composition law
\be
M_{2n,\,1\to 1}=\begin{cases}M_{n+r,\,1\to 1} M_{n-r,\,1\to 1} + M^{(2)}_{n+r,\,1\to 3} M^{(2)}_{n-r,\,1\to 3} & (n+r: \mbox{ even}) \\
                                            M^{(1)}_{n+r,\,1\to 2} M^{(1)}_{n-r,\,1\to 2} + M^{(1)}_{n+r,\,1\to 3} M^{(1)}_{n-r,\,1\to 3} & (n+r: \mbox{ odd}). \end{cases}
\ee                                            

\subsection{For $\lambda_1=0$, $\lambda_2> 0$ : }
Compared with the case of $\lambda_1>0$ and $\lambda_2=0$, 
in this case there is a drastic change in the behavior of the ground states as the maximum height we can reach in a given path is only 2. 
This is due to the fact that we are no longer allowed to come down by $\ket{\left(x_{3,1}\right)}$ (or $\ket{\left(x_{3, 2}\right)}$) 
once we go up by $\ket{\left(x_{2,3}\right)}$ (or $\ket{\left(x_{1,3}\right)}$). 
Thus we lose two of the equivalence classes, $\{12\}$ and $\{21\}$ reducing the GSD from 4 to 2. 

This choice of the parameters changes the recursion relations of the generating functions for computing the normalization of the state in the zero height case to 
\begin{eqnarray}
N_{1\to 1}(x) - 1 & = & x^2\left(N_{2\to 2}(x)N_{1\to 1}(x) + N_{1\to 1}(x)\right), \\
N_{2\to 2}(x) - 1 & = & x^2N_{2\to 2}(x).
\end{eqnarray}
The solution of the last equation, $N_{2\to 2}(x) = \frac{1}{1-x^2}$, can be understood from the observation that there is just one state in the $\{22\}$ equivalence class 
and this is given by $\prod\limits_{j=1}^{n} \ket{\left(x_{2,3}\right)_{2j-1}}\otimes\ket{\left(x_{3,2}\right)_{2j}}$ for length-$2n$. 
Solving for $N_{1\to 1}(x)$ leads to 
\begin{equation}
N_{1\to 1}(x) = \frac{1-x^2}{1-3x^2+x^4}
\end{equation}
which is an even function making the odd coefficients, $N_{2n+1,\,1\to 1}=0$. This is expected for the Dyck paths which make sense only for an even number of paths. 

The leading order behavior of the coefficients of these two terms is given by
\begin{equation}
N_{n,\,1\to 1} \sim \frac{1+(-1)^n}{2}\frac{1}{\sqrt{5}}\left(\frac{\sqrt{5}+1}{2}\right)^{n+1}, \qquad N_{n,\,2\to 2} = \frac{1+(-1)^n}{2}.
\end{equation}

The generating functions for the normalizations in the case of nonzero heights are obtained as 
\begin{eqnarray}
N^{(1)}_{1\to 2}(x) & = & \frac{x}{1-3x^2+x^4}, \\ 
N^{(1)}_{1\to 3}(x) & = & \frac{x(1-x^2)}{1-3x^2+x^4}, \\ 
N^{(2)}_{1\to 3}(x) & = & \frac{x^2}{1-3x^2+x^4}, \\ 
N^{(1)}_{2\to 3}(x) & = & \frac{x}{1-x^2}.
\end{eqnarray}
The rest are zero due to the height restriction as noted earlier. 

The leading order behavior of these coefficients are given by
\begin{eqnarray}
N^{(1)}_{n,\,1\to 2} & \sim & \frac{1-(-1)^n}{2} \frac{1}{\sqrt{5}} \left(\frac{\sqrt{5}+1}{2}\right)^{n+1} , \label{282} \\   
N^{(1)}_{n,\,1\to 3} & \sim & \frac{1-(-1)^n}{2} \frac{1}{\sqrt{5}} \left(\frac{\sqrt{5}+1}{2}\right)^{n},  \label{283} \\  
N^{(2)}_{n,\,1\to 3} & \sim & \frac{1+(-1)^n}{2} \frac{1}{\sqrt{5}} \left(\frac{\sqrt{5}+1}{2}\right)^{n}, \label{284} \\    
N^{(1)}_{2\to 3} & = & \frac{1-(-1)^n}{2} \label{285}.
\end{eqnarray}

\section{Entanglement entropy of the ground states}
\label{sec:EEGS}

\subsection{For $\lambda_1> 0$, $\lambda_2 = 0$ :}

The normalized ground state $\{11\}$ for the system of length $2n$ is expressed as 
\be
\ket{P_{2n,\, 1\to 1}} =\frac{1}{\sqrt{N_{2n,\,1\to 1}}} \,\sum_{w\in P_{2n,\,1\to 1}} \ket{w},
\label{GS11}
\ee
where $w$ runs over paths in $P_{2n,\,1\to 1}$. We split the system of length $2n$ into two subsystems A and B of length $n+r$ and $n-r$, respectively. 
Consider paths in $P_{2n,\, 1\to 1}$ that reach the point $(n,h)$, with the arrow index $a$. 
The paths belonging to A are $P^{(h)}_{n+r,\,1\to a} \equiv P^{(0\to h)}_{n+r,\,1 \to a}$ to denote that it starts at height 0 and ends at height $h$. And for B we have the reversed paths of 
$P^{(h)}_{n-r,\,1\to a}$ denoted by $P^{(h\to 0)}_{n-r,\, a\to 1}$. 
Their corresponding normalized states are expressed by 
\be
\ket{P^{(0\to h)}_{n+r,\,1\to a}} = \frac{1}{\sqrt{N^{(h)}_{n+r,\,1\to a}}} \,\sum_{w\in P^{(0\to h)}_{n+r,\,1\to a}} \ket{w},
\quad 
\ket{P^{(h\to 0)}_{n-r,\, a\to 1}} = \frac{1}{\sqrt{N^{(h)}_{n-r,\,1\to a}}} \,\sum_{w\in P^{(h\to 0)}_{n-r,\,a\to 1}} \ket{w},
\label{GShi1}
\ee
respectively. 

The Schmidt decomposition of (\ref{GS11}) leads to the following formula for the EE
\be
\ket{P_{2n,\, 1\to 1}} =\sum_{h\geq 0}\sum_{a=1}^3\sqrt{p^{(h)}_{n+r, n-r,\, 1\to a\to1}}\, \ket{P^{(0\to h)}_{n+r,\,1\to a}} \otimes \ket{P^{(h\to 0)}_{n-r,\, a\to 1}} , 
\label{GS11decomp}
\ee
where $p^{(h)}_{n+r, n-r,\, 1\to a\to1}\equiv \frac{N^{(h)}_{n+r,\,1\to a} N^{(h)}_{n-r,\,1\to a}}{N_{2n,\,1\to 1}}$ 
satisfies
\be
\sum_{h\geq 0}\sum_{a=1}^3 p^{(h)}_{n+r, n-r\, 1\to a\to1}=1
\ee
 due to (\ref{composition}) with $a=c=1$.

The reduced density matrix for the subsystem A takes the diagonal form as 
\be
\rho_{A, \, 1\to 1} = \Tr_B \ket{P_{2n,\, 1\to 1}} \,\bra {P_{2n,\, 1\to 1}} 
=\sum_{h\geq 0}\sum_{a=1}^3 p^{(h)}_{n+r, n-r,\, 1\to a\to1} \ket{P^{(0\to h)}_{n+r,\,1\to a}} \, \bra {P^{(0\to h)}_{n+r,\,1\to a}}, 
\ee
from which the EE reads 
\be
S_{A,\, 1\to 1}= -\sum_{a=1}^3\sum_{h\geq 0} p^{(h)}_{n+r, n-r,\, 1\to a\to1}\, \ln p^{(h)}_{n+r, n-r,\, 1\to a\to1}.
\label{EE11}
\ee

By using (\ref{Nij_asym}), (\ref{Nh11_asym}), (\ref{Nh12_asym}) and (\ref{Nh13_asym}),  
we find the logarithmic violation of the area law showing quantum criticality:
\be
S_{A, \, 1\to 1}=\frac12\ln \frac{(n+r)(n-r)}{n} + \frac12\ln\frac{\pi}{4} +\gamma -\frac12 +\mbox{(terms vanishing as $n\to \infty$)}
\label{EE11f}
\ee
with $\gamma$ being the Euler constant. 
We can see that this behavior including the constant term coincides with the case of the uncolored Fredkin spin chain~\cite{dyck}~\footnote{
For the uncolored Fredkin model, the number of paths of the Dyck walks $N^{(h)}_n$ in (\ref{Nhn}) is relevant, and its asymptotic form is given by (\ref{Nhn_asym}). 
In terms of $p^{(h)}_{n+r, n-r}\equiv \frac{N^{(h)}_{n+r} N^{(h)}_{n-r}}{N^{(0)}_{2n}}$, the EE is expressed as $S_A=-\sum_{h\geq 0} p^{(h)}_{n+r, n-r}\, \ln p^{(h)}_{n+r, n-r}$.
}
. 
For other ground states in equivalence classes $\{12\}$, $\{21\}$ and $\{22\}$, we obtain the same result.

\subsection{For $\lambda_1=\lambda_2=0$ :}
\label{sec:EE00}

In this case, we compute the EE of the ground state corresponding to paths equivalent to $\left(x_{1,2}x_{2,1}\right)^n$. 
The EE denoted by $\bar{S}_{A,\,1\to 1}$ takes the same form as (\ref{EE11}), where 
\be
p^{(h)}_{n+r, n-r,\, 1\to a \to 1} =\frac{M^{(h)}_{n+r,\,1\to a}M^{(h)}_{n-r,\,1\to a}}{M_{2n,\,1\to 1}}
\ee
with $M^{(0)}_{n\pm r,\,1\to a}\equiv M_{n\pm r,\,1\to a}$. 
From (\ref{M2nasymp}), (\ref{M112asymp}), (\ref{M113asymp}) and (\ref{M213asymp}), we find 
\bea
& & p^{(0)}_{n+r, n-r,\,1 \to 1\to 1}\sim \frac{1+(-1)^{n+r}}{2}\,\frac{\sqrt{5}+1}{2\sqrt{5}}, \qquad
p^{(1)}_{n+r, n-r,\,1 \to 2\to 1}\sim \frac{1-(-1)^{n+r}}{2}\,\frac{\sqrt{5}+1}{2\sqrt{5}}, \nn \\
& & p^{(1)}_{n+r, n-r,\,1 \to 3\to 1}\sim \frac{1-(-1)^{n+r}}{2}\,\frac{\sqrt{5}-1}{2\sqrt{5}}, \qquad
p^{(2)}_{n+r, n-r,\,1 \to 3\to 1}\sim \frac{1+(-1)^{n+r}}{2}\,\frac{\sqrt{5}-1}{2\sqrt{5}}, \nn \\
\eea
and all the others vanish. 
The results immediately lead to the expression of the EE: 
\be
\bar{S}_{A,\, 1\to 1}=\frac{1}{\sqrt{5}}\,\ln\frac{\sqrt{5}-1}{2} +\frac12\,\ln 5+(\mbox{terms vanishing as $n\to \infty$}), 
\ee
which exhibits the area law behavior. 

\subsection{For $\lambda_1=0$, $\lambda_2>0$ :}

In this case, the EE $S_{A,\,1\to 1}$ takes the same form as (\ref{EE11}), with 
$p^{(h)}_{n+r, n-r,\, 1\to a \to 1}$ computed using (\ref{282})-(\ref{285}). The non-zero terms are given by
\begin{eqnarray}
& & p^{(0)}_{n+r, n-r,\,1 \to 1\to 1}\sim \frac{1+(-1)^{n+r}}{2}\,\frac{\sqrt{5}+1}{2\sqrt{5}},\qquad p^{(1)}_{n+r, n-r,\,1 \to 2\to 1}\sim \frac{1-(-1)^{n+r}}{2}\,\frac{\sqrt{5}+1}{2\sqrt{5}} 
 \nn \\
& & p^{(1)}_{n+r, n-r,\,1 \to 3\to 1}\sim \frac{1-(-1)^{n+r}}{2}\,\frac{\sqrt{5}-1}{2\sqrt{5}}, \qquad
p^{(2)}_{n+r, n-r,\,1 \to 3\to 1}\sim \frac{1+(-1)^{n+r}}{2}\,\frac{\sqrt{5}-1}{2\sqrt{5}}. \nn \\
\end{eqnarray}

Thus, the resulting EE can be easily computed as
\begin{equation}
S_{A,\, 1\to 1}  =   \frac{1}{\sqrt{5}} \ln\frac{\sqrt{5} - 1}{2} + \frac{1}{2} \ln 5  
  +  (\mbox{terms vanishing as $n\to \infty$}). 
\end{equation}
For the $\{ 22\}$ equivalence class, the ground state is a product state corresponding to $\left(x_{2,3}x_{3,2}\right)^n$, whose EE vanishes ($S_{A,\,2\to 2}=0$).

\subsection{Quantum phase transitions :}

\begin{figure}[h]
\captionsetup{width=0.8\textwidth}
\centering
\includegraphics[scale = 0.8]{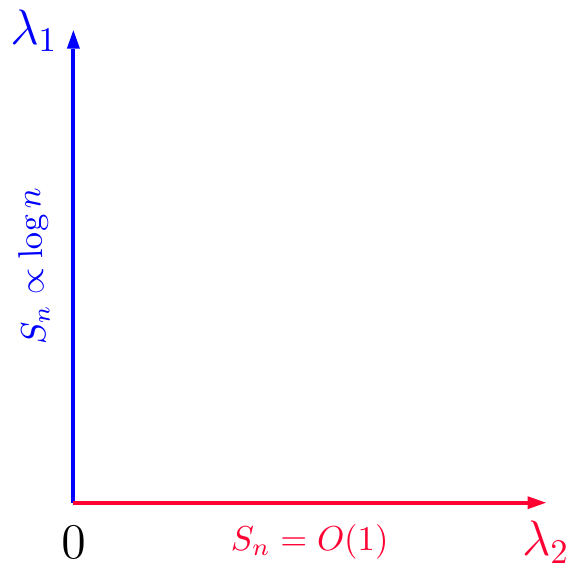}
\caption{\small The phase diagram for the modified Fredkin chain Hamiltonian in (\ref{hs31}). 
Area law is seen for values $\lambda_2 \geq 0$ with $\lambda_1=0$, and the logarithmic violation of the area law is for $\lambda_1>0$, $\lambda_2=0$. 
We do not discuss the case where $\lambda_1, \lambda_2 > 0$.} 
\label{phase_fredkin} 
\end{figure}

The phase diagram for the Hamiltonian (\ref{hs31}) in terms of the tunable parameters $\lambda_1$ and $\lambda_2$ is shown in Fig.~\ref{phase_fredkin}. 
We discuss three cases, 1) $\lambda_1> 0, \, \lambda_2 = 0$, 2) $\lambda_1 =  \lambda_2 = 0$ and 3) $\lambda_1 = 0, \,\lambda_2 > 0$. 
The latter two cases have ground states obeying the area law but are not the same phase as the GSD is not the same in them.

\section{Excitations due to disconnections}
\label{sec:excitations}
In this section, we discuss excited states corresponding to disconnected paths and their localization properties 
in the modified Fredkin spin chain (\ref{hs31}) with $\lambda_1>0$ and $\lambda_2=0$. 
Such states gain positive energy only from $H_{bulk, \,disconnected}$ (\ref{H_bulk_disconnected}). 

\subsection{Excited states with one disconnection}
\begin{figure}[h]
\captionsetup{width=0.8\textwidth}
\centering
\includegraphics[scale = 0.8]{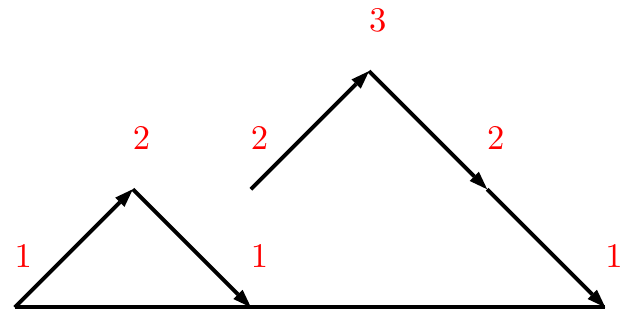}
\caption{\small The disconnected path (\ref{5_dc_path}) is depicted by following the convention.} 
\label{fig:5path1dc} 
\end{figure}

Before discussing general cases, let us start with an example of a length-5 path with one disconnection: 
\be
x_{1,2}x_{2,1} \textcolor{red}{|} x_{2,3}x_{3,2}x_{2,1}, 
\label{5_dc_path}
\ee
which consists of the two connected components $x_{1,2}x_{2,1}$ and $x_{2,3}x_{3,2}x_{2,1}$. The red vertical line denotes the disconnection. 
In drawing this configuration, {\it a priori} there is no way to determine the relative height between the points across the 
disconnection. 
Here and in what follows, we take a convention that the initial and final points of paths of the total length $n$ are $(0,0)$ and $(n,0)$, respectively. 
Then, (\ref{5_dc_path}) is depicted in Fig.~\ref{fig:5path1dc}. 

As a result of the local equivalence moves, we have 
\be
\left(x_{1,2}x_{2,1}+x_{1,3}x_{3,1}\right) \textcolor{red}{|}\left(x_{2,3}x_{3,2}x_{2,1}+x_{2,1}x_{1,2}x_{2,1}+x_{2,1}x_{1,3}x_{3,1}\right)
=P^{(0\to 0)}_{2,\,1\to 1} \textcolor{red}{|} P^{(1\to 0)}_{3,\,2\to 1}.
\label{excitation_ex1}
\ee
Note that the equivalence moves keep the disconnection intact, and just affect each of the connected components. 
The moves preserve the arrow indices of the two ends and the relative height between 
the end points of the connected components. 
The state corresponding to (\ref{excitation_ex1}), 
$\ket{P^{(0\to 0)}_{2,\,1\to 1}} \otimes \ket{P^{(1\to 0)}_{3,\,2\to 1}}$, 
costs energy only at the disconnected point, and thus has eigenvalue 1. 

In general, an excited state corresponding to length-$n$ paths starting (ending) at the arrow index $a$ ($c$) and possessing one disconnection at the site $i$ ($0<i<n$) is written as 
\be
\ket{P^{(0\to h_i)}_{i,\,a\to b_i}}\otimes \ket{P^{(h_i'\to 0)}_{n-i,\,b_i'\to c}} \qquad (b_i\neq b_i'), 
\label{1K}
\ee
where the disconnection requires $b_i\neq b_i'$. The heights $h_i$ and $h_i'$ take nonnegative integers, and they can take the same value. 
The local equivalence moves act only on the connected components of (\ref{1K}). 
The energy of the state is 1 due to the single disconnection.

\subsection{Excited states with two disconnections}
\begin{figure}[h]
\captionsetup{width=0.8\textwidth}
\centering
\includegraphics[scale = 0.8]{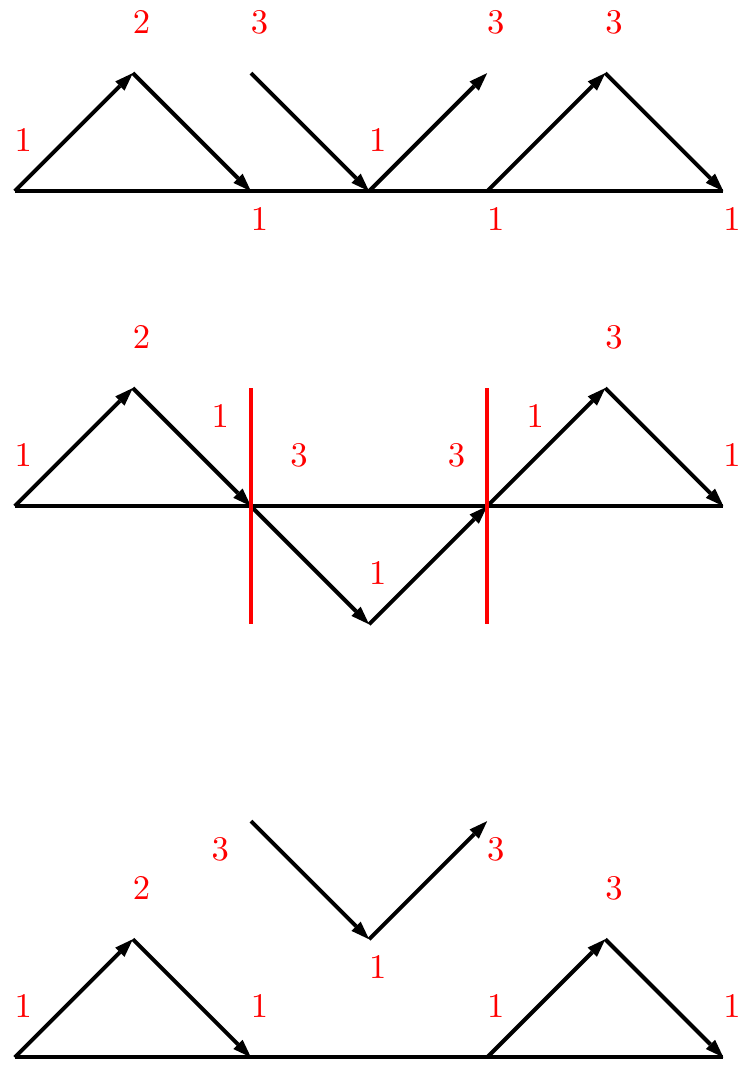}
\caption{\small These three figures represent the same path (\ref{6_dc_path}). In the middle figure, the red vertical lines are 
added in order to clarify the disconnections.} 
\label{fig:6path2dc} 
\end{figure}

We first consider an example of a length-6 path with two disconnections:
\be
x_{1,2}x_{2,1} \textcolor{red}{|} x_{3,1}x_{1,3} \textcolor{red}{|} x_{1,3}x_{3,1}. 
\label{6_dc_path}
\ee
In drawing this, the heights of the first and third connected components are fixed by the above convention, but the 
height of the second one is not. In general, the heights of the connected components are not fixed except for those of 
the first and the last ones. We will discuss without giving a definite prescription for the issue. 
So, note that the path (\ref{6_dc_path}) can be depicted in different ways as shown in Fig.~\ref{fig:6path2dc}.  

Applying the equivalence moves to (\ref{6_dc_path}) amounts to the paths  
\be
P^{(0\to 0)}_{2,\,1\to 1} \textcolor{red}{|} \left( x_{3,1}x_{1,3}+x_{3,2}x_{2,3}\right) \textcolor{red}{|} P^{(0\to 0)}_{2,\,1\to 1}.
\ee
The corresponding state 
\be
\ket{P^{(0\to 0)}_{2,\,1\to 1}}\otimes 
\frac{1}{\sqrt{2}} \left\{ \ket{\left(x_{3,1}\right)_3, \left(x_{1,3}\right)_4}+\ket{\left(x_{3,2}\right)_3, \left(x_{2,3}\right)_4}\right)\}
\otimes \ket{P^{(0\to 0)}_{2,\,1\to 1}}
\ee
has eigenvalue 2 due to the two disconnections. 

We can write a general excited state corresponding to length-$n$ paths starting (ending) at the arrow index $a$ ($c$) and possessing two disconnections 
at the sites $i$ and $j$ ($0<i<j<n$) as 
\be
\ket{P^{(0\to h_i)}_{i,\,a\to b_i}}\otimes \ket{\bar{P}^{(h_i'\to h_j)}_{j-i,\,b_i'\to b_j}} \otimes\ket{P^{(h_j'\to 0)}_{n-j,\,b_j'\to c}} \qquad (b_i\neq b_i',\, b_j\neq b_j'), 
\ee
where the arrow indices should be $b_i\neq b_i'$ and $b_j\neq b_j'$ for the disconnections. 
We take all the heights $h_i$, $h_i'$, $h_j$ and $h_j'$ nonnegative integers.  
The length-$(j-i)$ paths $\bar{P}^{(h_i'\to h_j)}_{j-i,\,b_i'\to b_j}$, appearing in the middle, start (end) at 
the height $h_i'$ ($h_j$) and the arrow index $b_i'$ ($b_j$), but not restricted to the region $y\geq 0$. 
Note that the first and the last connected components $P^{(0\to h_i)}_{i,\,a\to b_i}$ and $P^{(h_j'\to 0)}_{n-j,\,b_j'\to c}$ have the restriction due to the boundary terms $H_{left}$ and $H_{right}$. 
The definition of the state $\ket{\bar{P}^{(h_i'\to h_j)}_{j-i,\,b_i'\to b_j}}$ is similar to (\ref{GShi1}), normalized by the number of the connected paths $\bar{P}^{(h_i'\to h_j)}_{j-i,\,b_i'\to b_j}$. 
This excited state has energy 2.

\subsection{Excited states with $K$ disconnections}
A general excited state possessing $K$ disconnections at the sites $i_1, \cdots, i_K$  ($0<i_1<\cdots <i_K<n$) is expressed as 
\begin{align}
&\ket{P^{(0\to h_{i_1})}_{i_1,\,a\to b_{i_1}}}\otimes \ket{\bar{P}^{(h'_{i_1}\to h_{i_2})}_{i_2-i_1,\,b'_{i_1}\to b_{i_2}}} \otimes\cdots \otimes \ket{\bar{P}^{(h'_{i_{K-1}}\to h_{i_K})}_{i_K-i_{K-1},\,b'_{i_{K-1}}\to b_{i_K}}} 
\otimes \ket{P^{(h'_{i_K}\to 0)}_{n-i_K,\,b'_{i_K}\to c}} \nn \\
& (b_{i_1}\neq b'_{i_1},\cdots, b_{i_K}\neq b'_{i_K}), 
\label{excitationK}
\end{align}
where $K$ disconnections mean $b_{i_1}\neq b'_{i_1},\cdots, b_{i_K}\neq b'_{i_K}$.  
All the heights $h_{i_1}, h'_{i_1}, \cdots, h_{i_K}, h'_{i_K}$ are nonnegative integers.  
The connected components except the first and last are not restricted to $y\geq0$. 
The energy eigenvalue of the state is $K$, which is equal to the number of the disconnections. 

\begin{figure}[h]
\captionsetup{width=0.8\textwidth}
\centering
\includegraphics[scale = 0.8]{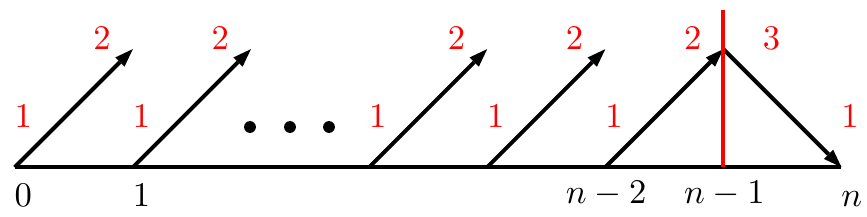}
\caption{\small The totally disconnected path (\ref{totallydisc}). The red vertical line denotes the disconnection.} 
\label{fig:totallydisc} 
\end{figure}

The $K=n-1$ case yields totally disconnected paths. The corresponding states are not entangled at all, and gain the maximal energy $n-1$ due to the disconnections. 
An example of such paths and its corresponding state are 
\bea
 & & x_{1,2}\cdots x_{1,2}\textcolor{red}{|} x_{3,1}, 
 \label{totallydisc}
 \\
& & \ket{\left(x_{1,2}\right)_1}\otimes\cdots\otimes \ket{\left(x_{1,2}\right)_{n-1}}\otimes \ket{\left(x_{3,1}\right)_n}
\eea
which is depicted in Fig.~\ref{fig:totallydisc}.  

\subsection{Excited states with localization}
\subsubsection{Highly excited localized states}
\begin{figure}[h!]
\captionsetup{width=0.8\textwidth}
\begin{center}
		\includegraphics[scale=0.8]{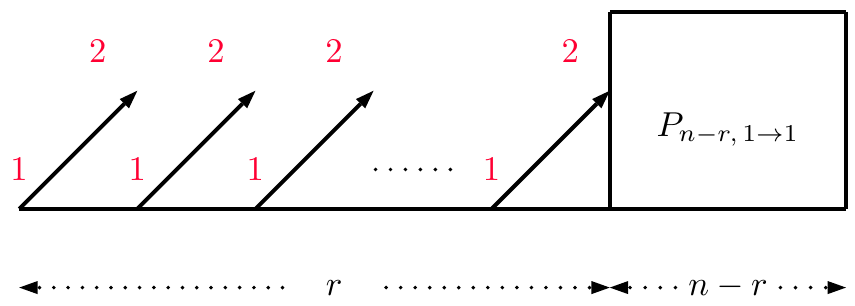} 
	\caption{\small A highly excited state where $r$ links are totally disconnected and the remaining $n-r$ links are fully connected to the state $\{11\}$.}
\label{fig:he}
\end{center}
\end{figure}

For excited states with localization, we first consider highly excited states for the Hamiltonian $H_F$, 
which are partially connected with a large region being totally disconnected and the complement being fully connected. One such state is given by
\begin{equation}
\ket{he} = \left(\otimes_{i=1}^r~\ket{\left(x_{1,2}\right)_i}\right)~\otimes~\ket{P_{n-r,\, 1\to 1}},
\label{he_state}
\end{equation}
with $r$ of the links totally disconnected and the remaining $n-r$ links being the fully connected $\{11\}$ state (a ground state for just the $n-r$ links). 
Such a state has the energy eigenvalue $r$. When $r$ is large, the state is highly excited as $he$ in the l.h.s. denotes. This is shown in Fig.~\ref{fig:he}. 

We expect this state to be localized that is the connected part does not interact with the totally disconnected region, 
implying no spread of information between these two regions. 
To show this, we study the connected time correlations of local operators acting on these two regions 
\be
\bra{he}\theta_{i}(t)\theta_j(0)\ket{he} - \bra{he}\theta_{i}(t)\ket{he}\bra{he}\theta_j(0)\ket{he},
\label{correlation}
\ee
for local operators at the site $i$ and $j$: $\theta_i$ and $\theta_j$. 
We take $j$ to index a link inside the connected part of the state on the $n-r$ links and the index $i$ to label a link in the first $r$ steps of the chain. 
We will show that this correlation is zero for the operators acting on the local Hilbert spaces in this system. 
These operators include the flip operators 
\begin{equation} \label{flip}
\theta_i^{(a_1,b_1;\,a_2,b_2)}(0) \equiv \ket{x_{a_1, b_1}}_i\bra{x_{a_2, b_2}}, \qquad a_1\neq a_2~\textrm{or}~b_1\neq b_2,
\end{equation}
with the arrow indices $a_1, a_2, b_1, b_2 \in \{1,2,3\}$, and the diagonal operators 
\begin{equation}\label{diag}
\theta_i^{(a,b)}(0) \equiv \ket{x_{a, b}}_i\bra{x_{a, b}}, \qquad a, b\in\{1,2,3\}~\textrm{and}~a\neq b.
\end{equation}

When $\theta_j(0)$ is a flip operator, 
it is easy to note that $\theta_j^{(a_1,b_1;\,a_2,b_2)}(0)\ket{he}$ disconnects the fully connected state on the $n-r$ links and produces a partially connected eigenstate with a single disconnection, 
which is not modified by further applying the operator (\ref{flip}) or (\ref{diag}) with time evolution $\theta_i(t) = e^{iH_Ft}\theta_i(0)e^{-iH_Ft}$. 
Since this partially connected excitation does not have an overlap with $\ket{he}$ as the $\ket{he}$ has a connected component on $n-r$ links, 
we can see that the first and second terms in the correlation (\ref{correlation}) separately vanish. Thus, (\ref{correlation}) is zero. 

For cases that operators $\theta_m(0)$ ($m = i, j$) are diagonal in the local Hilbert space, 
we work with 
\be
\theta_i^{\{\kappa\}}(0) = \sum_{a, b=1, a\neq b}^3~\kappa_{ab}\,\theta_i^{(a, b)}(0), 
\qquad 
\theta_j^{\{\kappa'\}}(0) = \sum_{a, b=1, a\neq b}^3~\kappa'_{ab}\,\theta_i^{(a, b)}(0),
\ee
where $\kappa_{ab}$, $\kappa'_{ab}$ are coefficients. 
The computation goes as follows
\begin{equation}
\theta_j^{\{\kappa'\}}(0) \ket{he} = \sum_{\alpha}~b_{\alpha} \ket{\alpha},
\end{equation}
where $\ket{\alpha}$ are partially connected energy eigenstates of $H_F$ that have the same form as $\ket{he}$. 
Namely, the first $r$ links are occupied by the $\ket{x_{1, 2}}$ state and the remaining $n-r$ links are connected components in the $\{11\}$ equivalence class. 
One of the states $\ket{\alpha}$ also coincides with $\ket{he}$, and the coefficients $b_\alpha$ depend on $\kappa'_{ab}$. 
Now let us apply $\theta_i^{\{\kappa\}}(t) = e^{iH_Ft}\theta_i^{\{\kappa\}}(0)e^{-iH_Ft}$,
\begin{equation}
e^{iH_Ft}\theta_i^{\{\kappa\}}(0)e^{-iH_Ft} \sum_{\alpha}~b_{\alpha} \ket{\alpha} = \kappa_{12} \sum_{\alpha}~b_{\alpha} \ket{\alpha},
\end{equation} 
as the first $r$ links are totally disconnected and composed only by $\ket{x_{1,2}}$.  
Thus we have 
\begin{equation}
\bra{he}\theta_{i}^{\{\kappa\}}(t)\theta_j^{\{\kappa'\}}(0)\ket{he} = \kappa_{12} b_{he}.
\end{equation}
On the other hand, it is easy to see 
\begin{equation}
\bra{he}\theta_{i}^{\{\kappa\}}(t)\ket{he} = \kappa_{12},
\qquad 
\bra{he}\theta_{j}^{\{\kappa'\}}(0)\ket{he} = b_{he},
\end{equation}
which leads to the vanishing connected correlator: 
\be
\bra{he}\theta_{i}^{\{\kappa\}}(t)\theta_j^{\{\kappa'\}}(0)\ket{he} - \bra{he}\theta_{i}^{\{\kappa\}}(t)\ket{he}\bra{he}\theta_{j}^{\{\kappa'\}}(0)\ket{he} = 0. 
\ee 

Similarly, the correlator vanishes for cases of $\theta_i$ being flipping and $\theta_j$ diagonal. 
 
General local operators acting on the local Hilbert space of the modified Fredkin chain can be constructed by linear combinations of the flip and diagonal operators 
(\ref{flip}) and (\ref{diag}) as 
\be
\theta_m^{\{\kappa\}}(0)=\sum_{a_1\neq a_2\, \textrm{or}\,b_1\neq b_2}\kappa_{(a_1,b_1;\,a_2,b_2)}\theta_m^{(a_1,b_1;\,a_2,b_2)}(0)
+\sum_{a, b=1, a\neq b}^3~\kappa_{ab}\,\theta_m^{(a, b)}(0)
\label{general_localop}
\ee
with $\kappa_{(a_1,b_1;\,a_2,b_2)}$ and $\kappa_{ab}$ coefficients. 
Thus, the connected correlators (\ref{correlation}) for all the local operators vanish, 
implying there is no spread of information between the disconnected and the connected regions of $\ket{he}$. 

Not strictly local operators $\hat{\cO}_i$ ($\hat{\cO}_j$) with some locality range $\Delta$ around $i$ ($\Delta'$ around $j$) are expressed by products of (\ref{general_localop}): 
\be
\hat{\cO}_{i,\,\Delta}(0) \equiv \prod_{\ell=i-\Delta}^{i+\Delta} \theta_\ell^{\{\kappa_\ell\}}(0), \qquad
\hat{\cO}_{j,\,\Delta'}(0) \equiv \prod_{\ell=j-\Delta'}^{j+\Delta'} \theta_\ell^{\{\kappa_\ell\}}(0). 
\label{localop_Delta}
\ee
We can again show that the connected correlator vanishes when $i+\Delta<r<j-\Delta'$ ($\hat{\cO}_{i,\,\Delta}(0)$ acting only on the totally disconnected links, and 
$\hat{\cO}_{j,\,\Delta'}(0)$ only on the connected $n-r$ links).

\subsubsection{Generalization}
\begin{figure}[h!]
\captionsetup{width=0.8\textwidth}
\begin{center}
		\includegraphics[scale=0.8]{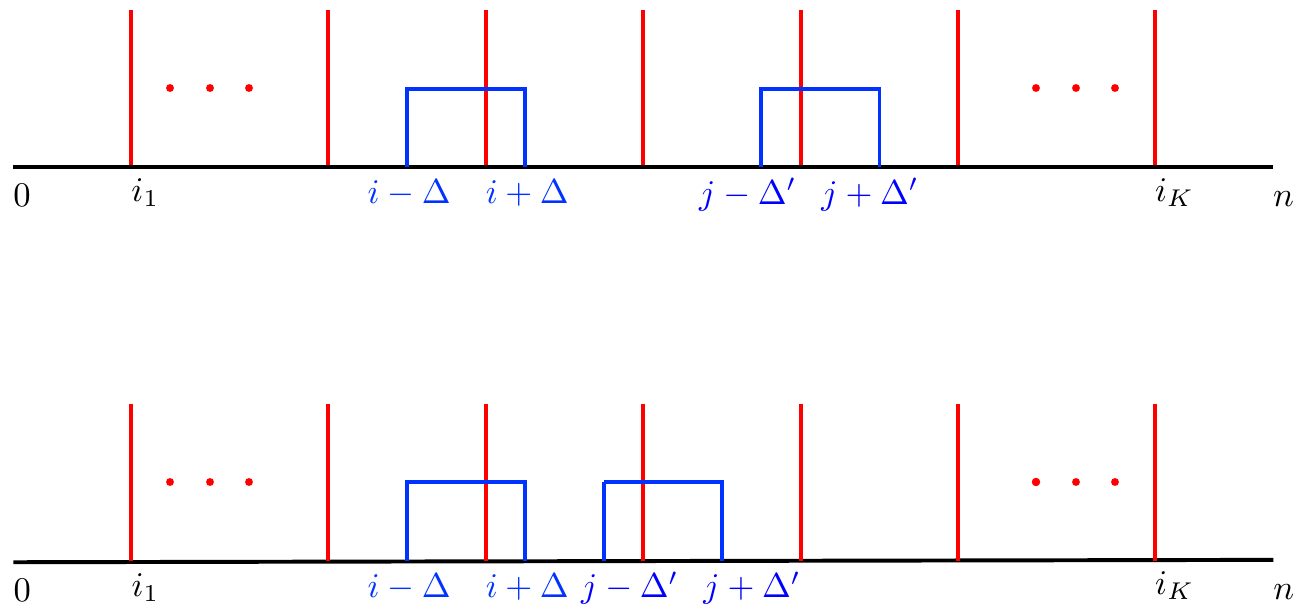} 
	\caption{\small In both of the two panels, the red vertical lines show disconnections and the blue lines represent the locality ranges of the operators $\hat{\cO}_{i,\,\Delta}(0)$ and $\hat{\cO}_{j,\,\Delta'}(0)$. 
	Upper panel: two connected components that cover the range $[i-\Delta,\,i+\Delta]$ are distinct from those that cover $[j-\Delta',\, j+\Delta']$, yielding the vanishing connected correlation function. 
	Lower panel: two connected components covering $[i-\Delta,\,i+\Delta]$ have an overlap with those covering $[j-\Delta',\, j+\Delta']$. The correlator does not vanish in general.}
\label{fig:localizationK}
\end{center}
\end{figure}

Here we discuss the localization for the excited states with $K$ disconnections (\ref{excitationK}). 
The point in the previous cases is that the states $\left(\otimes_{i=1}^r~\ket{\left(x_{1,2}\right)_i}\right)$ and $\ket{P_{n-r,\, 1\to 1}}$ in (\ref{he_state}) undergo independent time evolutions even after the action of the local operators, 
from which we can easily see 
\be
\bra{he}\hat{\cO}_{i,\,\Delta}(t)\hat{\cO}_{j,\,\Delta'}(0)\ket{he} = \bra{he}\hat{\cO}_{i,\,\Delta}(t)\ket{he}\bra{he}\hat{\cO}_{j,\,\Delta'}(0)\ket{he}, 
\ee 
meaning that the connected correlation function is zero. This point of view immediately allows us to show the localization for the states (\ref{excitationK}). 
{\it The two point connected correlation function of $\hat{\cO}_{i,\,\Delta} (t)$ and $\hat{\cO}_{j,\,\Delta'}(0)$ on the state (\ref{excitationK}) vanishes, when the connected component including the range $[i-\Delta,\, i+\Delta]$ 
does not overlap with $[j-\Delta',\, j+\Delta']$.} 
In cases that a single connected component does not cover the range $[i-\Delta,\, i+\Delta]$ or  $[j-\Delta',\, j+\Delta']$, we consider the minimal connected components that can cover the range. 
When the minimal connected components for the range $[i-\Delta,\, i+\Delta]$ do not overlap those for $[j-\Delta',\, j+\Delta']$, the two-point connected correlator vanishes. 
Fig.~\ref{fig:localizationK} shows examples for the case giving the vanishing connected correlator (upper panel) and the case when the correlator is nonvanishing (lower panel).    
In the former case, information is confined in the minimal connected components. 
 
In this subsection, we have localized states that are (highly) excited in the spectrum of the modified Fredkin chain. 
It is also worth noting that no disorder was introduced into the system to create these localized many-body states, which are in contrast to the ordinary cases of Anderson localization~\cite{anderson} 
or many-body localization~\cite{basko}~\footnote{
For reviews, see \cite{MBLrev1,MBLrev2,MBLrev3}.
}. 
This is consistent with the fact that we have conserved operators in this system given by 
\be
\cO_{i,\,i+1}^{(ab,cd)}\equiv \ket{\left(x_{a, b}\right)_i,\,\left(x_{c, d}\right)_{i+1}}\bra{\left(x_{a, b}\right)_i,\,\left(x_{c, d}\right)_{i+1}} \qquad (a\neq b,\, b\neq c,\, c\neq d), 
\label{LIOM}
\ee
where $i$ runs from 1 to $n-1$. These operators are the summand of $H_{bulk,\,disconnected}$ (\ref{H_bulk_disconnected}). 
It is easy to see that they commute with the Hamiltonian $H_F$ and mutually commute, as noted earlier. 
The number of the operators (\ref{LIOM}) is $24(n-1)$, whereas the total degrees of freedom of the system is $6^n$.

\section{Discussions}
\label{sec:discussions}
In this paper, we have constructed an extended model of the Fredkin spin chain based on the SIS $\cS^3_1$. 
It is similar to the previous work for the modified Motzkin spin chains~\cite{FSPP}, although the connection to the SIS is less and partial 
because of the absence of the elements corresponding to flat steps $x_{a,a}$ ($a=1,2,3$). In the parameter space $(\lambda_1, \lambda_2)$ with 
nonnegative $\lambda_1$ and $\lambda_2$ satisfying $\lambda_1\lambda_2=0$, we have found the three phases, I) $\lambda_1>0, \lambda_2=0$, 
II) $\lambda_1=\lambda_2=0$ and III) $\lambda_1=0, \lambda_2>0$. In I), there are four degenerate ground states, which exhibit a logarithmic violation 
of the area law of the EE, signaling quantum criticality. 
Although the EE obeys the area law for both of II) and III), GSD grows with the length of the system in II), whereas GSD is fixed to 2 in III).  
Quantum phase transitions occur at the boundaries of the phases. 
As a feature of the extended model, there are excited states due to disconnections with respect to arrow indices, which exhibit localization phenomena without 
any disorder, in contrast to the usual case of Anderson and many-body localization. 
It is clear that the same localization takes place in the modified Motzkin model, although such excitations are not discussed in~\cite{FSPP}.

In the forthcoming paper~\cite{PPFSVK2}, we investigate the extended model with color degrees of freedom, which is based on the SIS $\cS^3_2$ and its generalizations. 
Similar to what was seen in the modified Motzkin case~\cite{FSPP}, the model has stronger entanglement, i.e. the EE of the ground states is 
proportional to the square root of the volume. The localization properties of excited states are also discussed.  

Deformations of the Motzkin and Fredkin spin chains discussed in refs.~\cite{i1,i2,i3,i4} change the equal weighted sum over the paths in the ground state to a weighted sum, which realizes 
the extensive EE proportional to the volume. It is interesting to realize the volume law or other scaling properties of EE in our setting associated to 
SISs.




\end{document}